\documentclass[preprint,12pt,sort&compress]{elsarticle}

\usepackage{amsmath,amssymb}

\usepackage{graphicx}
\usepackage[left=2cm,right=2cm,top=2cm,bottom=2cm]{geometry}
\usepackage{xcolor}

\usepackage[bookmarks=true,colorlinks=true,citecolor=blue,linkcolor=blue,urlcolor=magenta]{hyperref}

\begin{document}

\begin{frontmatter}

\title{Coronal heating problem solution by means\\of axion origin photons}

\author[addr1]{Vitaliy~D.~Rusov\corref{cor1}}
\cortext[cor1]{Corresponding author e-mail: siiis@te.net.ua}

\author[addr1]{Igor~V.~Sharph}

\author[addr1]{Vladimir~P.~Smolyar}

\author[addr2]{Maxim~V.~Eingorn}

\author[addr3]{Margarita~E.~Beglaryan}

\address[addr1]{
	Department of Theoretical and Experimental Nuclear Physics,\\
	Odessa National Polytechnic University, Odessa, Ukraine
}

\address[addr2]{
	CREST and NASA Research Centers, North Carolina Central University,\\
	Durham, North Carolina, U.S.A.
}

\address[addr3]{
	Department of Computer Technology and Applied Mathematics,\\
	Kuban State University, Krasnodar, Russia
}

\date{October 29, 2020}

\begin{abstract}

In this paper we advocate for the idea that two seemingly unrelated mysteries with almost 90 year history -- the nature of dark matter and the million-degree solar corona -- may be but two sides of the same coin -- the axions of dark matter born in the core of the Sun and photons of axion origin in the million-degree solar corona, whose modulations are controlled by the anticorrelated modulation of the asymmetric dark matter (ADM) density in the solar interior.

It is shown that the photons of axion origin, that are born in the
almost empty magnetic flux tubes (with $B \sim 10^7 ~G$) near the tachocline and then pass through the
photosphere to the corona, are the result of the solar corona heating
variations, and thus, the Sun luminosity variations. Since the spectrum of 
the incident photons of axion origin is modulated by the frequency dependence of the
cross-section, then, first, the energy distribution of the emitted axions is
far from being a blackbody spectrum, and second, for a typical solar spectrum,
the maximum of the differential axion flux occurs at the average axion energy is $\langle E_a / T \rangle \approx 4.4$ 
(Raffelt,1986).
This means that the average energy of the photon of axion origin can generate a
temperature of the order of $T_a \sim 1.11 \cdot 10^7 ~K$ under certain
conditions of coronal substances, which is close to the temperature
$T_{core} \sim 1.55 \cdot 10^7 ~K$ of the Sun core.

As a result, the free energy accumulated by the photons of axion origin in a
magnetic field by means of degraded spectra due to multiple Compton scattering,
is quickly released and converted into heat and plasma motion with a
temperature of $\sim 4 \cdot 10^6~K$ at maximum and $\sim 1.5 \cdot 10^6~K$ at
minimum of solar luminosity.

Since the photons of axion origin are the result of the Sun luminosity variations, then, unlike the self-excited dynamo, an unexpected but simple question arises: is there a dark matter chronometer hidden deep in the Sun core?

A unique result of our model is the fact that the periods, velocities and modulations of S-stars are the fundamental indicator of the modulation of the ADM halo density in the fundamental plane of the Galaxy center, which closely correlates with the density modulation of the baryon matter near the SMBH. If the modulations of the ADM halo at the GC lead to modulations of the ADM density on the surface of the Sun (through vertical density waves from the disk to the solar neighborhood), then there is an ``experimental'' anticorrelation
identity between the indicators, e.g.
the modulation of the ADM density in the solar interior and the number of sunspots.
Therefore, this is also true for the modulation of the ADM density in the solar interior, which is directly related to the identical periods of S-star cycles and the sunspot cycles.
\end{abstract}


\end{frontmatter}

\section{Introduction}

A hypothetical pseudoscalar particle called axion is predicted by the theory
related to solving the CP-invariance violation problem in QCD. The most
important parameter determining the axion properties is the energy scale $f_a$
of the so-called U(1) Peccei-Quinn symmetry violation. It determines both the
axion mass and the strength of its coupling to fermions and gauge bosons
including photons. However, in spite of the numerous direct experiments, axions
have not been discovered so far. Meanwhile, these experiments together with the
astrophysical and cosmological limitations leave a rather narrow band for the
permissible parameters of invisible axion (e.g.
$10^{-6} eV \leqslant m_a \leqslant 10^{-2} eV$~\citep{ref01,ref02}), which is
also a well-motivated cold dark matter (CDM) candidate in this mass region
\citep{ref01,ref02}.

Let us give some implications and extractions from the photon-axion
oscillations theory which describes the process of the photon conversion into
an axion and back under the constant magnetic field $B$ of the length $L$. It
is easy to show \citep{ref05,Raffelt-Stodolsky1988,Mirizzi2005,Hochmuth2007}
that in the case of the negligible photon absorption coefficient
($\Gamma _{\gamma} \to 0$) and axions decay rate ($\Gamma _{a} \to 0$) the
conversion probability is

\begin{equation}
P_{a \rightarrow \gamma} = \left( \Delta_{a \gamma}L \right)^2 \sin ^2 \left( \frac{ \Delta_{osc}L}{2} \right) \Big/ \left( \frac{ \Delta_{osc}L}{2}
\right)^2 \label{eq01}\, ,
\end{equation}

\noindent
where the oscillation wavenumber $\Delta_{osc}$ is given by

\begin{equation}
\Delta_{osc}^2 = \left( \Delta_{pl} + \Delta_{Q,\perp} - \Delta_{a} \right)^2 + 4 \Delta_{a \gamma} ^2
\label{eq02}
\end{equation}

\noindent
while the mixing parameter $\Delta _{a \gamma}$, the axion-mass parameter
$\Delta_{a}$, the refraction parameter $\Delta_{pl}$ and the QED dispersion
parameter $\Delta_{Q,\perp}$ may be represented by the following expressions:

\begin{equation}
\Delta _{a \gamma} = \frac{g_{a \gamma} B}{2} = 540 \left( \frac{g_{a \gamma}}{10^{-10} GeV^{-1}} \right) \left( \frac{B}{1 G} \right) ~~ pc^{-1}\, ,
\label{eq03}
\end{equation}

\begin{equation}
\Delta _{a} = \frac{m_a^2}{2 E_a} = 7.8 \cdot 10^{-11} \left( \frac{m_a}{10^{-7} eV} \right)^2 \left( \frac{10^{19} eV}{E_a} \right) ~~ pc^{-1}\, ,
\label{eq04}
\end{equation}

\begin{equation}
\Delta _{pl} = \frac{\omega ^2 _{pl}}{2 E_a} = 1.1 \cdot 10^{-6} \left( \frac{n_e}{10^{11} cm^{-3}} \right) \left( \frac{10^{19} eV}{E_a} \right) ~~ pc^{-1},
\label{eq05}
\end{equation}

\begin{equation}
\Delta _{Q,\perp} = \frac{m_{\gamma, \perp}^2}{2 E_a} .
\label{eq06}
\end{equation}

Here $g_{a \gamma}$ is the constant of axion coupling to photons; $B$ is the
transverse magnetic field; $m_a$ and $E_a$ are the axion mass and energy;
$\omega ^2 _{pl} = 4 \pi \alpha n_e / m_e$ is an effective photon mass in terms
of the plasma frequency if the process does not take place in vacuum, $n_e$ is
the electron density, $\alpha$ is the fine-structure constant, $m_e$ is the
electron mass; $m_{\gamma, \perp}^2$ is the effective mass square of the
transverse photon which arises due to interaction with the external magnetic
field.

The conversion probability (\ref{eq01}) is energy-independent, when
$2 \Delta _{a \gamma} \approx \Delta_{osc}$, i.e.

\begin{equation}
P_{a \rightarrow \gamma} \cong \sin^2 \left( \Delta _{a \gamma} L \right)\, ,
\label{eq07}
\end{equation}

\noindent
or whenever the oscillatory term in (\ref{eq01}) is small
($\Delta_{osc} L / 2 \to 0$), implying the limiting coherent behavior
\begin{equation}
P_{a \rightarrow \gamma} \cong \left( \frac{g_{a \gamma} B L}{2} \right)^2\, .
\label{eq08}
\end{equation}

It is worth noting that the oscillation length corresponding to (\ref{eq07})
reads
\begin{equation}
L_{osc} = \frac{\pi}{\Delta_{a \gamma}} = \frac{2 \pi}{g_{a \gamma} B} \cong 5.8 \cdot 10^{-3}
\left( \frac{10^{-10} GeV^{-1}}{g_{a \gamma}} \right)
\left( \frac{1G}{B} \right)  ~pc
\label{eq13}
\end{equation}

\noindent assuming a purely transverse field. In the case of the appropriate
size $L$ of the region a complete transition between photons and axions is
possible.

From now on we are interested in the energy-independent case (\ref{eq07}) or
(\ref{eq08}) which plays the key role in determination of the parameters for
the axion mechanism of Sun luminosity variations (the axion coupling constant
to photons $g_{a\gamma}$, the transverse magnetic field $B$ of length $L$ and
the axion mass $m_a$).

To estimate the hadron axion-photon coupling constant, we focus on the
conversion probability

\begin{equation}
P_{a \rightarrow \gamma} = \frac{1}{4} \left( g_{a \gamma} B_{MS} L_{MS} \right)^2 \sim 1 , ~ ~ L_{MS}\equiv L_{osc}
\label{eq3.31}
\end{equation}

\noindent
where the complete conversion between photons and axions is possible by means
of estimating the axion coupling constant to photons.

We showed a solution to the well-known problem of the Parker-Biermann cooling
effect, which is determined by the nature of strong toroidal magnetic fields
in tachocline (by the thermomagnetic Ettingshausen-Nernst effect
(\ref{appendix-a}))
and provides a physical basis for the channeling
of photons (of axion origin), which are born on the boundary of the tachocline,
inside the practically empty magnetic flux tubes
(MFTs; \ref{appendix-b}).
The manifestation of the
tachocline itself is (in contrast to the self-excited dynamo) the result of the
fundamental holographic principle of quantum gravity
(\ref{appendix-c}).

It should be remembered that, on the one hand, for the Coulomb field of
a charged particle in the solar core, the conversion is best seen as a process
of electron-nuclear collisions $\gamma + (e,Ze) \rightarrow (e,Ze) + a$, called
incoherent Primakoff effect \cite{Raffelt1986}. On the other hand, reverse
conversion is considered as a process of axions converting into $\gamma$-quanta
in a magnetic field $a + \vec{B} \rightarrow \vec{B} + \gamma$ and called the inverse
coherent Primakoff effect~\cite{Raffelt1986}.

\begin{figure*}[tbp!]
  \begin{center}
    \begin{minipage}[h]{0.44\linewidth}
      \includegraphics[width=\linewidth]{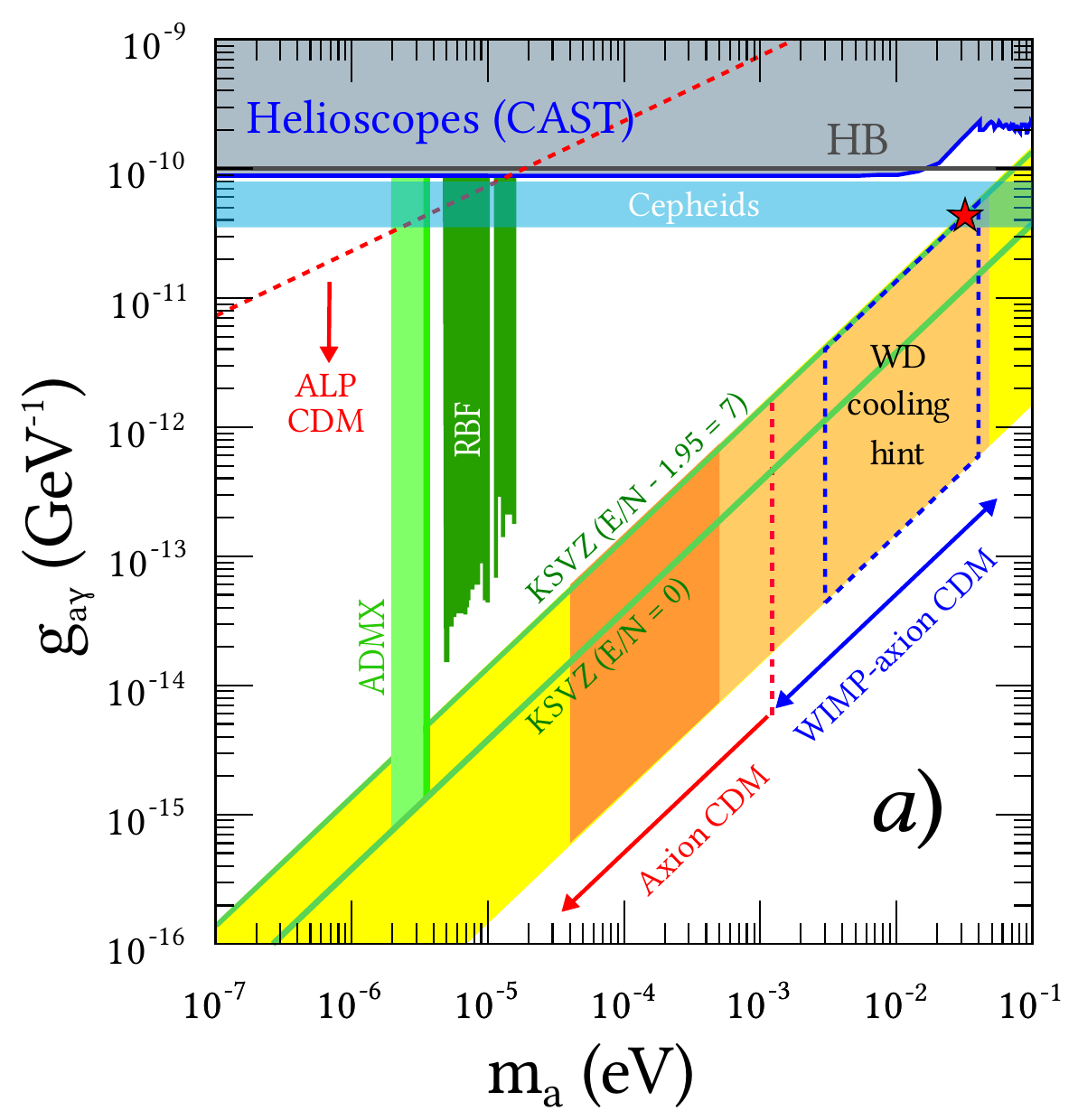}
    \end{minipage}
    \hfill
    \begin{minipage}[h]{0.53\linewidth}
      \includegraphics[width=\linewidth]{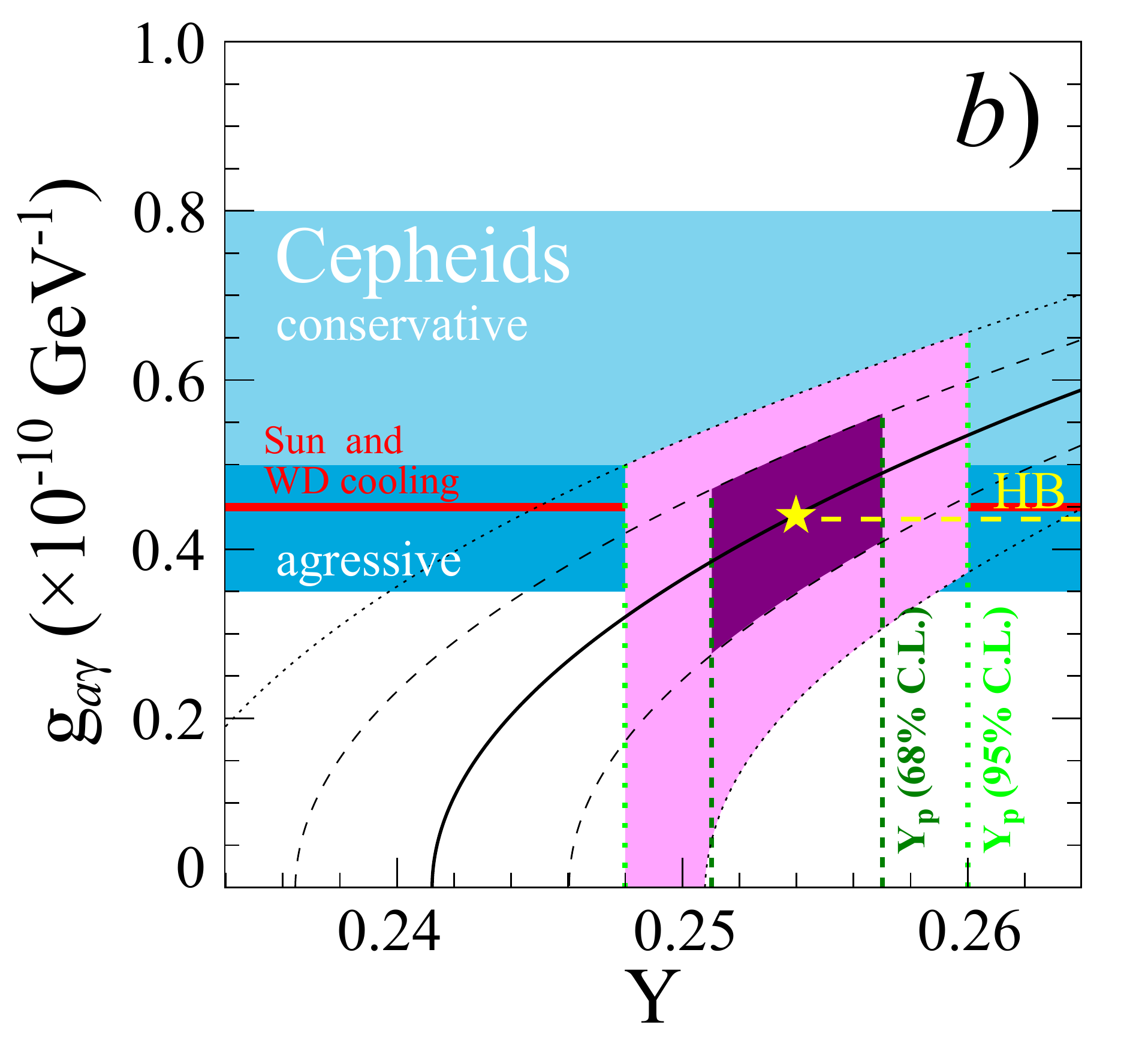}
    \end{minipage}
  \end{center}
\caption{\textbf{(a)} Summary of astrophysical, cosmological and laboratory
constraints on axions and ALPs. Comprehensive axion/ALP
parameter space, highlighting two main front lines of direct detection
experiments: helioscopes (CAST~\citep{Andriamonje2007,ref72,CAST2011,Arik2013}) and
haloscopes (ADMX~\citep{ref50} and RBF~\citep{ref51}). The astrophysical bounds
from horizontal branch and massive stars are labeled ``HB''~\citep{ref02} and
``Cepheids''~\citep{Carosi2013}, respectively. The QCD motivated models
(KSVZ~\citep{ref46,ref46a} and DFSZ~\citep{ref47,Dine1981}) for axions lay in
the yellow diagonal band. The orange parts of the band correspond to
cosmologically interesting axion models: models in the ``classical axion
window'' possibly composing the totality of DM (labeled ``Axion CDM'') or a
fraction of it (``WIMP-axion CDM''~\citep{Baer2011}). For more generic ALPs,
practically all allowed space up to the red dashed line may contain valid ALP
CDM models~\citep{Arias2012}. The region of axion masses invoked in the white dwarf
cooling anomaly is shown by the blue dashed line~\citep{Irastorza2013}. The red
star marks the values of the axion mass $m_a \sim 3.2 \cdot 10^{-2} eV$ and the
axion-photon coupling constant $g_{a\gamma} \sim 4.4 \cdot 10^{-11} GeV^{-1}$
chosen in the present paper on the basis of the suggested relation between the
axion mechanisms of the Sun and the white dwarf luminosity variations.
\newline
\textbf{(b)} $R$ parameter constraints on $Y$ and $g_{a \gamma}$ (adopted from
\cite{Ayala2014}). The dark purple area delimits the 68\%~C.L. for $Y$ and
$R_{th}$ (see Eq.~(1) in \cite{Ayala2014}). The resulting bound on the axion
($g_{10} = g_{a \gamma \gamma}/(10^{-10} ~GeV^{-1})$) is somewhere between
rather conservative $0.5 < g_{10} \leqslant 0.8$ and most aggressive $0.35 <
g_{10} \leqslant 0.5$ \citep{Friedland2013}. The red line marks the value of
the axion-photon coupling constant $g_{a \gamma} \sim 4.4 \cdot 10^{-11}
~GeV^{-1}$ chosen in the present paper.
The blue shaded area represents the bounds from Cepheids
observation. The yellow star corresponds to $Y$=0.254 and the bounds from HB
lifetime (yellow dashed line).}
\label{fig05}
\end{figure*}

According to \ref{appendix-a},
the $B_{MS} \sim 4100$~T magnetic field in the overshoot tachocline and the
Parker-Biermann cooling effect can produce the O-loops with the horizontal
magnetic field $B_{MS} \approx B(0.72 R_{Sun}) \sim 3600$~T stretching for about
$L_{MS} \sim 1.28 \cdot 10^4 ~km$, (see e.g. Eq.~(9) in \cite{RusovArxiv2019}),
and surrounded by virtually zero internal gas pressure of the magnetic tube
(see Fig.~6a,b in \cite{RusovArxiv2019})
Since $P_{a \rightarrow \gamma} \sim 1$, we obtain the following parameters of
the hadron axion (see Fig.~\ref{fig05}).

\begin{equation}
g_{a \gamma} \sim 4.4 \cdot 10^{-11} ~ GeV^{-1}, ~~~ m_a \sim 3.2 \cdot 10^{-2} ~eV.
\label{eq3.30}
\end{equation}

\noindent
where the axion mass $m_a$ can be expressed through the properties of $\pi^0$-meson~\cite{Bardeen1978}:

\begin{equation}
m_a = \frac{m_\pi f_\pi}{f_a} \left( \frac{z}{(1+z)(1+z+\omega)} \right)^{1/2} ,
\end{equation}

\noindent
where $f_a$ is the energy scale~\cite{PecceiQuinn1977,Weinberg1978,Wilczek1978},
$m_\pi$ and $f_\pi$ are the pion mass and decay constant respectively, while
$z = m_u / m_d$ and $\omega = m_u / m_s$ are the quark mass ratios. The value
of $f_a$ can be easily obtained from the axion coupling constant to photons 
$g_{a \gamma}$ (Eq.~(\ref{eq3.31})) under the condition 
$P_{a \rightarrow \gamma} \sim 1$.

It is very important that in our work the theoretical estimate for the fraction
of the axion luminosity $L_a$ in the total luminosity of the Sun $L_{Sun}$
\citep{Andriamonje2007} with respect to (\ref{eq3.30}) is

\begin{equation}
\frac{L_a}{L_{Sun}} = 1.85 \cdot 10 ^{-3} \left( 
\frac{g_{a \gamma}}{10^{-10} GeV^{-1}} \right)^2 \sim 3.6 \cdot 10^{-4} .
\label{eq3.32}
\end{equation}

Thus, it is shown that the hypothesis about the possibility for the solar
axions born in the core of the Sun to be efficiently converted back into
$\gamma$-quanta in the magnetic field of the magnetic steps of the O-loop
(above the solar overshoot tachocline) is relevant. The almost empty magnetic
flux tubes with $B \sim 4.1 \cdot 10^7 ~G$
(\ref{appendix-a})
 is the
physical base for the almost zero gas pressure inside the tube, and 
consequently, the channeling of the photons of axion origin without absorption
and scattering. As a result, the variations of the magnetic field in the solar
tachocline are the direct cause of the converted $\gamma$-quanta intensity
variations. The latter, in their turn, may be the cause of the overall solar
luminosity variations known as the active and quiet Sun phases.

Considering the above remarks, we believe that it is necessary to calculate the
effect of axion-originated photons in active and quiet phases of the Sun on the
solar corona. 

To do this, in Sec.~\ref{sec-corona-heating-problem} we first briefly examine
the current problem of the corona heating. 

In \mbox{Sec.~\ref{sec-nanoflares-spectrum}} we discuss the part of the axion and
nanoflares spectra in the total solar spectrum during active and quiet phases
of the Sun. In \mbox{Sec.~\ref{sec-corona-heating-scenario}} we outline a scenario of
corona heating by means of axion origin photons. In 
\mbox{Sec.~\ref{sec-corona-heating-solution}} we suggest a solution to the corona
heating problem by means the photons of axion origin.

In Sec.~\ref{sec-dark-matter} we obtain an indirect proof of the connection
between the periods of S-stars near the black hole, and the sunspot cycles.
Finally, in Sec.~\ref{sec-summary} we give a brief summary of this work.

\section{Corona heating problem}
\label{sec-corona-heating-problem}

An interesting problem arises here related to the heating of the practically empty magnetic tube 
(see Figs.~6a and 11 in \cite{RusovArxiv2019}) 
and heating of the solar corona, which has long been unresolved (see \cite{DeMortel2015,DeMortel2016}). There are many assumptions about the unusually high temperature in the corona (see e.g. $T_{average} \approx (1.5 - 3) \cdot 10^6 ~K$ in \cite{Gudel2009} and \cite{Kariyappa2011,Cirtain2013,Peter2014,Reale2014,Laming2015,Peter2015,Morgan2017}) compared to the chromosphere and photosphere. It is known that energy comes from the underlying layers, including, in particular, the photosphere and chromosphere. Here are just some of the elements, possibly involved in the heating of the corona: 
magneto-acoustic and Alfv\'{e}n waves 
(see \cite{Alfven1947,Schatzman1949,Schatzman1962,Parker1964,Callebaut1994,
Jess2009,Jess2016,Ballegooijen2011,Ballegooijen2014,Arregui2015,Morton2016,
Laming2017,Vigeesh2017}), 
magnetic reconnections (see e.g. \cite{Shibata1999,Watanabe2011,Archontis2014,Xue2016,Sun2015,Huang2018}), 
nano-flares (see e.g. \cite{Aulanier2013,Testa2014,Brosius2014,Cargill2015,Klimchuk2015}), 
Ellerman bombs (see e.g. \cite{Watanabe2011,Vissers2013,Vissers2015,Libbrecht2017}). 
It is considered (see e.g. \cite{Jess2009}) that the possible mechanism of corona heating is the same as for the chromosphere: convective cells in the form of granulation rising from the depth of the Sun and appearing in the photosphere (see e.g. \cite{Ballegooijen2011,Ballegooijen2014,Dudik2014,Cranmer2015}) lead to a local imbalance in the gas, which leads to the propagation of magneto-acoustic and Alfv\'{e}n waves (see e.g. Fig.~13 in \cite{Laming2015}) moving in different directions. In this case, a chaotic change in the density, temperature and velocity of the substance in which these waves propagate leads to a change in the speed, frequency and amplitude of the magneto-acoustic waves and can be so large that the gas becomes supersonic. Shock waves appear (see \cite{Solanki2003,Grib2014,Santamaria2016}), which lead to the heating of the gas and, as a consequence, to the heating of the corona.

On the other hand, we believe that the main effective mechanism for heating the solar corona is  the emission of axions from the solar core with an energy spectrum with the maximum of about 3~keV and the average of 4.2~keV. These axions are supposed to convert into soft X-rays in very strong transverse magnetic field of an almost empty tube at the base of the convective zone 
(\mbox{\ref{appendix-b}}). Eventually we suppose that the X-rays,
through the axion-photon conversions in the magnetic O-loop near the
tachocline, channel along the ``cool'' region of the Parker-Biermann magnetic
tube (Fig.~\mbox{\ref{fig-lampochka}}) and effectively supply the necessary
photons of axion origin ``channeling'' in an almost empty magnetic tube to the
photosphere while the convective heat transfer is heavily suppressed 
(\mbox{\ref{appendix-b}}).
As a result, X-rays, passing through the photosphere at high speed and scattering in the Compton process, reach the transition region between the chromosphere and corona. The bulk of soft X-rays dissipates in the corona (see Fig.~\ref{fig-corona-compton}c,d).

The total power radiated by X-rays of axion origin is only about one millionth of the total solar luminosity,
so there is sufficient energy on the Sun to heat the corona.

It should be recalled here that the energy distribution of the emitted axions is far from being a blackbody spectrum because the spectrum of the incident photons is modulated by the frequency dependence of the cross-section. For the typical solar spectrum, the maximum of the differential flux of axions occurs at $E_a / T \approx 3.5$, whereas the average axion energy is $\langle E_a / T \rangle \approx 4.4$ \citep{Raffelt1986}. This means that the average energy of the photon of axion origin can generate a temperature of the order of $T_a \sim 1.11 \cdot 10^7 ~K$ under certain conditions of coronal substances \citep{Priest2000}, which is close to the temperature $T_{core} \sim 1.55 \cdot 10^7 ~K$ of the Sun core (see e.g. \cite{Fiorentini2001,Fiorentini2002,Bahcall1992,Bahcall1995}).

This raises an intriguing question about the paradoxical heating of the corona 
(see e.g. \cite{Edlen1943,Alfven1947,Parker1958,Gibson1973,Withbroe1977,
Parker1988,Klimchuk2006,Tomczyk2007,Aschwanden2007,Erdelyi2007,Golub2009,
Pontieu2011,Parnell2012,Reale2014,AschwandenEtAl2014,AschwandenEtAl2015,
AschwandenEtAl2016,AschwandenEtAl2017,Tan2014,DeMortel2015,Klimchuk2015,
Barnes2016,Morton2016}): how do the soft X-rays of axion origin heat the solar corona and flares  to a temperature of more than two and, correspondingly, three orders of magnitude higher than in the photosphere?

Below we show why the solar corona is so hot with the help of photons of axion origin.

\subsection{The complete solar spectrum of nanoflares and axions in the solar atmosphere}
\label{sec-nanoflares-spectrum}

\begin{figure*}
  \begin{center}
    \includegraphics[width=18cm]{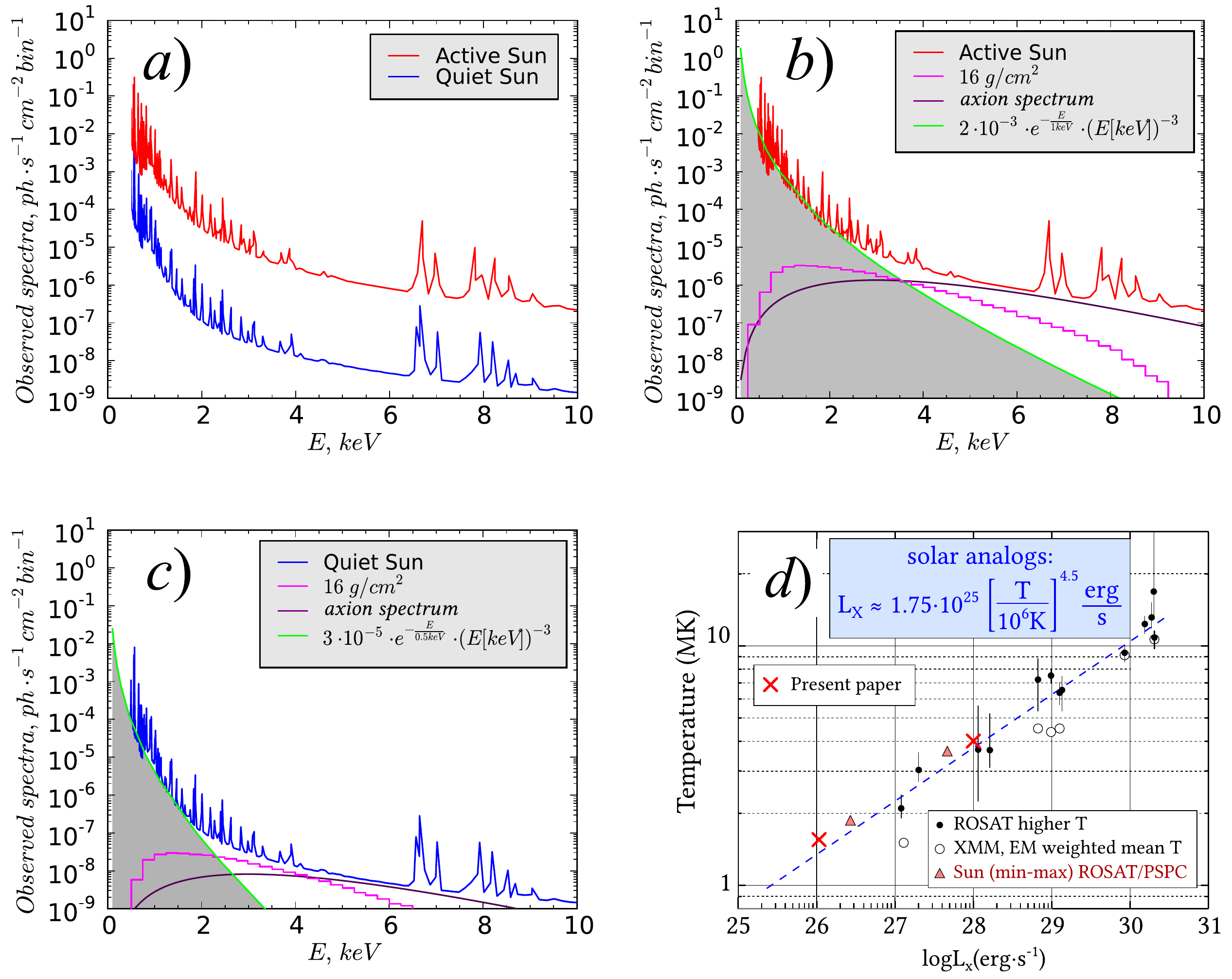}
  \end{center}

\caption{\textbf{(a)} Reconstructed solar photon spectrum in the 0.5\,-\,10~keV band from 
the active
Sun (red line) and quiet Sun (blue line) from accumulated observations
(spectral bin is 6.1~eV wide)~\citep{Peres2000}.
\newline
\textbf{(b)} Reconstructed solar photon spectrum fit in the active phase of the Sun by
the quasi-invariant soft part of the solar photon spectrum (grey shaded area;
see Eq.~(\ref{eq06-34})) and the spectrum (\ref{eq3.33}) degraded to
the Compton scattering for column density above the initial conversion place
of 16~$g / cm^2$ (see Fig.~10 in \cite{Zioutas2009}), and the
initial energy distribution (thin line in violet) is the converted solar
axion spectrum.
Here the inverse Compton effect was not taken into account for the range
8~keV - 10~keV.
 Note that the GEANT4 code photon threshold is at 1~keV 
(reconstructed solar photon spectra from \cite{Peres2000}).
\newline
\textbf{(c)} Similar curves for the quiet phase of the Sun (grey shaded area 
corresponds to \mbox{Eq.~(\ref{eq06-35})}).
\newline
\textbf{(d)} 
Corona temperature against the X-ray luminosity of the solar
analogs (see ROSAT, XMM-Newton, adopted from \cite{Gudel2004}). The data from
\textit{ROSAT/PSPC} are shown as
compared to the maximum and minimum of the solar cycle (see Sect.3.3 in
\cite{Peres2000}).}
\label{fig06}
\end{figure*}

From the axion mechanism point of view the solar spectra during
the active and quiet phases (i.e. during the maximum and minimum solar
activity) differ from each other by the soft part ($0.5~keV < E < 2.0~keV$),
where the power-law spectra of nanoflares prevail, or the hard part
($2~keV < E < 10~keV$) of the Compton spectrum. The latter being
produced by the photons of axion origin
ejected from the magnetic tubes into the photosphere (see Fig.~\ref{fig06} and Fig.~4 in
\cite{ChenF2015}).

A natural question arises at this point: ``What are the real parts of the
Compton spectrum of axion origin in the active and quiet phases of the Sun,
and do they agree with the experiment?'' Let us perform the
mentioned estimations being based on the known experimental results by ROSAT/PSPC,
where the Sun coronal X-ray spectra and the total luminosity during the
minimum and maximum of the solar coronal activity were obtained~\citep{Peres2000}.

Apparently, the solar photon spectrum below 10~keV of the active and quiet Sun
(Fig.~\ref{fig06}a) reconstructed from the accumulated ROSAT/PSPC observations
can be described by two components: the first component is the soft
($0.5~keV < E < 2.0~keV$) part of the spectrum where the power-law spectra of
nanoflares dominate, and the second is the harder ($2~keV < E < 10~keV$) part
of the spectrum which was degraded to the Compton scattering for column density
above the initial conversion place of 16~$g/cm^2$ (see Fig.~10 
in~\cite{Zioutas2009}), and the initial energy distribution is
the converted solar axion spectrum (Fig.~\ref{fig06}b,c).
The complete solar spectrum is:

\begin{align}
\left( \frac{d \Phi}{dE} \right)^{(*)} \approx
\left( \frac{d \Phi}{dE} \right)^{(*)}_{nanoflare} +
\left( \frac{d \Phi _{\gamma}}{dE} \right)^{(*)}_{axions} ,
\label{eq06-33}
\end{align}

\noindent where ${d \Phi}/{dE}$ is the observed solar spectra during the
active (red line in Fig.~\ref{fig06}a,b) and quiet (blue line in
Fig.~\ref{fig06}a,c) phases, $\left({d \Phi}/{dE} \right)_{nanoflare}$
represents the power-law spectra of the nanoflares at 0.5-2.0~keV,

\begin{equation}
\left( \frac{d \Phi}{dE} \right)_{nanoflare} \sim E^{-(1+\alpha)} e^{-E/E_0} ,
\label{eq06-33a}
\end{equation}

\noindent
where the power-law decay with the ``semi-heavy tail'' takes place in practice 
\citep{Lu1993} instead of the so-called power laws with heavy tails 
\citep{Lu1991,Lu1993} (see e.g. Figs.~3 and~6 in \cite{Uchaikin2013}). 
Consequently, the observed corona spectra ($0.5 ~keV < E < 2.0 ~keV $)
(shaded area in Fig.~\ref{fig06}b)

\begin{align}
& \left( \frac{d \Phi}{dE} \right)^{(active)}_{nanoflare} \sim
2 \cdot 10^{-3} \cdot (E~[keV])^{-3} \cdot \exp{\left(-\frac{E}{1 keV} \right)} 
\nonumber \\
&~~for~the~active~Sun
\label{eq06-34}
\end{align}

\noindent and (shaded area in Fig.~\ref{fig06}c)

\begin{align}
&\left( \frac{d \Phi}{dE} \right)^{(quiet)}_{nanoflare} \sim
3 \cdot 10^{-5} \cdot (E~[keV])^{-3} \cdot \exp{\left(-\frac{E}{0.5 keV} \right)} 
\nonumber \\
&~~for~the~quiet~Sun ;
\label{eq06-35}
\end{align}

\noindent $\left( {d \Phi _{\gamma}}/{dE} \right)_{axions}$ is the
reconstructed solar photon spectrum fit ($2 ~keV < E < 10 ~keV$) built up from
two spectra (see Fig.~\mbox{\ref{fig06}b,c} for the active and quiet phases of
the Sun, respectively).

As is known, this class of flare models (Eqs.~(\ref{eq06-34})
and~(\ref{eq06-35})) is based on the recent paradigm in statistical physics
known as self-organized criticality
\citep{Bak1987,Bak1988,Bak1989,Bak1996,Aschwanden2011}. The basic idea is that
the flares are a result of an ``avalanche'' of small-scale magnetic
reconnection events cascading \citep{Lu1993,Charbonneau2001,Aschwanden2014} 
through the highly intense coronal magnetic structure \citep{Shibata2011} driven
at the critical state by the accidental photospheric movements of its magnetic
footprints. Such models thus provide the natural and computationally convenient
basis for the study of Parker's hypothesis of the coronal heating by nanoflares
\citep{Parker1988}.

Another significant fact discriminating the theory from practice, or rather 
giving a true understanding of the measurements against some theory, should be 
recalled here (see e.g. Eq.~(\ref{eq06-33a}); also see Eq.~(5) in \cite{Lu1993}). The 
nature of power laws is related to the strong connection between the consequent
events (this applies also to the ``catastrophes'', which  in turn give rise to
a spatial nonlocality related to the appropriate structure of the medium (see 
page 45 in \cite{Uchaikin2013})). As a result, the ``chain reaction'', i.e. the
avalanche-like growth of perturbation with more and more resource involved,
leads to the heavy-tailed distributions. On the other hand, obviously, none of 
the natural events may be characterized by the infinite values of the mean and 
variance. Therefore, the power laws like (\ref{eq06-33a}) are approximate and
must not hold for the very large arguments. It means that the power-law decay
of the probability density rather corresponds to the average asymptotics, and
the ``semi-heavy tails'' must be observed in practice instead.

\begin{figure*}[tbp]
\begin{center}
\includegraphics[width=18cm]{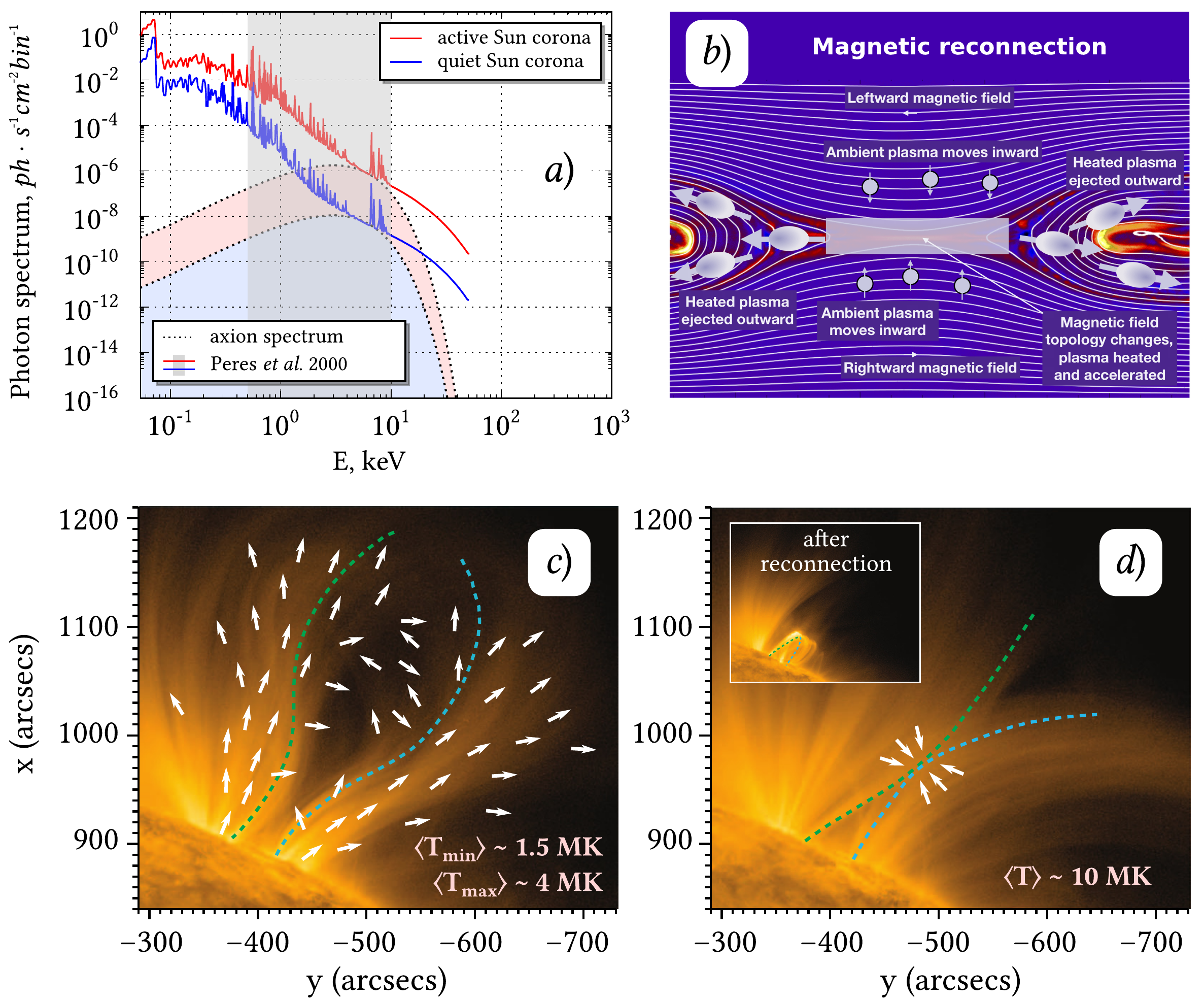}
\end{center}
\caption{\textbf{Top:} Synthesized photon spectra of the corona and 
nanoflares.
\textbf{(a)} The coronal total luminosity of the Sun (see Table~2 in~\cite{Peres2000}) in the ROSAT/PSPC band ranges from 
$\approx 2.7 \cdot 10^{26} ~erg/s$ at minimum to 
$\approx 4.7 \cdot 10^{27} ~erg/s$ at maximum.
To calculate the observed spectra of the corona, we used abbreviated data
($0.5~keV < E < 10.0~keV$) in~\cite{Peres2000} (Fig.~\ref{fig06}a,b,c).
\textbf{(b)} 
Simplified 2D schematic diagram of magnetic reconnection. Oppositely
directed magnetic fields (light blue lines) and ambient plasma (light blue
circles) move into the diffusion region (shaded box in the center), where
magnetic reconnection occurs. The plasma is heated and accelerated into jets
to the left and right (shaded blue ovals). Adopted from~\cite{Hesse2020}.
\textbf{Bottom:} The origin of magnetic flux and the associated dynamic coronal
phenomenon on the Sun \citep{Sun2015} and the appearance of the photon
fluxes of axion origin. The blue and green dashed curves show the selected
coronal loops representing the two lines of magnetic fields involved in the
process (c) and (d). The sequence of extreme ultraviolet images clearly shows
that two groups of oppositely directed and non-coplanar magnetic loops (c)
gradually approach each other (d),
causing a magnetic reconnection.
If the plasma attracts magnetic loops, which gradually approach each other,
causing magnetic reconnection, then nanoflares of two frequency types appear
(see Fig.~\ref{fig-nanoflares-reconnection}). In our case, when the 
plasma interacts with photons of axion origin, the magnetic reconnection of
the nanoflare is the result of low-frequency heating. As a result, rather rare
high-energy nanoflares occur with a low frequency, and the temperature rises
sharply to about 10~MK. On the other hand, when the plasma exists without
interacting with photons of axion origin, then frequent low-energy nanoflares
occur, and the temperature fluctuations are insignificant 
(see Fig.~\ref{fig-nanoflares-reconnection}).
}
\label{fig-corona-compton}
\end{figure*}

It is known \citep{Priest2000} that traditionally in the atmosphere of the Sun there are three types of eruptions, such as coronal mass ejections, prominence eruptions and eruptive flares, and they are considered bound and are the result of the same physical process. Coronal mass ejections (CMEs) are large-scale mass ejections and magnetic flux from the lower corona to the interplanetary space. It is believed that they should be created by the loss of equilibrium in the structures of the coronal magnetic plasma, which causes sharp changes in the magnetic topology. A typical CME carries approximately $10^{23} ~Mx$ of flux and $10^{13} ~kg$ of plasma into space \citep{Priest2000}. During the active phase of the solar cycle, CME can occur more often than once a day.
The intermittent appearance of a new magnetic flux from the convective zone (which originates from twist in flux tubes 
(see~\cite{Archontis2012,Schmieder2014,Pontieu2014}) in the corona is the most important process for the dynamic evolution of the coronal magnetic field 
\citep{Galsgaard2007,Fang2010,ArchontisHood2012,ArchontisHood2013}, in which the rearrangement of the intersection of closed coronal lines of force causes the accumulation of coronal field strength. When the stress exceeds a certain threshold, the stability of the magnetic field configuration is broken and erupts (see e.g. Fig.~\ref{fig-corona-compton}c; Fig.~2 in \cite{Sun2015}). This model is called a storage model \citep{Yamada2010}, although it is unfortunately known that the question of how magnetic fields rise from the tachocline to the convective zone of the Sun and exit through the photosphere and chromosphere into the corona has not yet been resolved (see e.g. \cite{Archontis2008,Bushby2012,Schmieder2015}).

Nevertheless, this plausible explanation is associated not only with the appearance of a magnetic flux, but also necessarily with the appearance of photons of axion origin from an empty magnetic tube 
(see Figs.~6a, 10a and 11 in \cite{RusovArxiv2019}), in which, according to our theoretical and experimental observations (see Fig.~\ref{fig-corona-compton}a), the simultaneous occurrence of a magnetic flux and the flux of photons of axion origin in the outer layers of the Sun is the main mechanism of the formation of sunspots and active regions, being the integral part of the solar cycle, and also of the high energy release in the corona and flares. The physical solution of this problem will be shown below.

First about the flares. The well-known scenario of magnetic 
reconnection\footnote{Let us recall the jocular but practically understandable
words of M.~Hesse and P.~A.~Cassak~\cite{Hesse2020} about the essence of
magnetic reconnection: ``There is an imprecise -- but useful -- analogy to
rubber bands that helps us picture magnetic reconnection. A loose rubber band
cannot hold a pile of pencils in place, but a stretched rubber band can. This
is because it takes energy to stretch the rubber band, and that energy can be
thought of as stored in the rubber band. The energy in the stretched rubber
band holds the pencils in place. The more you stretch a rubber band, the more
energy it stores. Eventually, if you stretch a rubber band too much, it breaks,
providing a painful lesson of how much energy it can hold!''}
suggests that the corona is heated by numerous small-scale energy release
events called nanoflares~\cite{Parker1988,Klimchuk2006,Klimchuk2015}.

According to~\cite{Klimchuk2015}, ``...'nanoflare' is a term that is
often used to describe an impulsive heating event. It was first coined by
Parker~\cite{Parker1988}, who envisioned a burst of magnetic
reconnection. The meaning of nanoflare has since evolved. Here, as in our
earlier work, we take it to mean an impulsive energy release on a small
cross-field spatial scale without regard to physical mechanism. It is a very
generic definition. Waves produce nanoflares (e.g.~\cite{Klimchuk2006}).
Much of the discussion in this paper concerns generic heating, including by
waves, but some of it deals specifically with magnetic reconnection.
The distinction should be obvious.''.

\begin{figure}[tb]
\includegraphics[width=\textwidth]{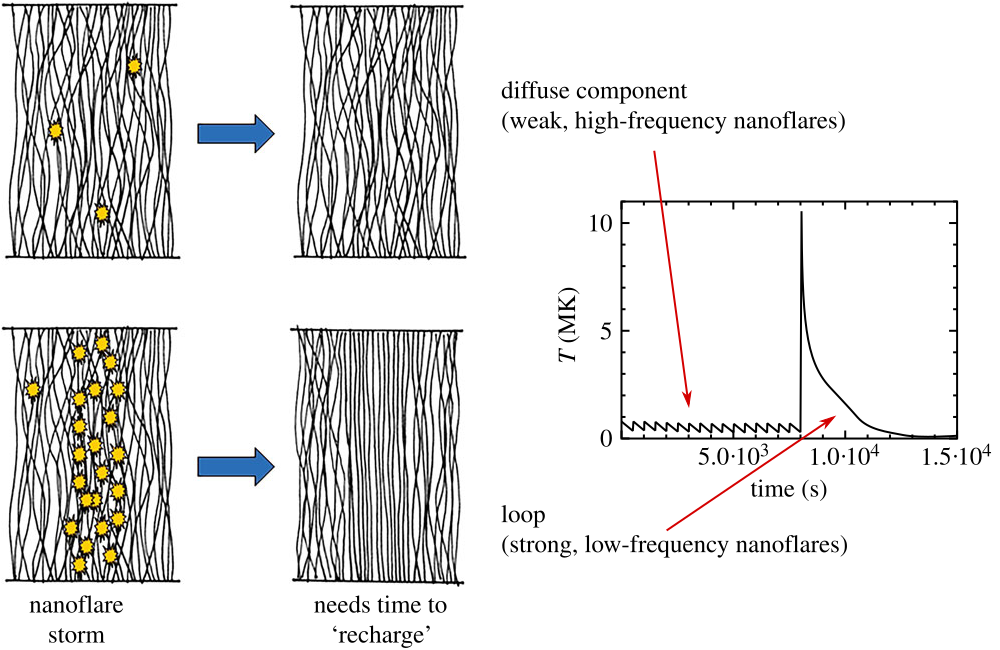}
\caption{A unifying picture that explains both the diffuse component of the
corona and distinct loops. Adapted from~\cite{Klimchuk2015}.}
\label{fig-nanoflares-reconnection}
\end{figure}

We are interested in heating with magnetic reconnection (see e.g. complementary
figures Fig.~1 in~\cite{Hesse2020} and Fig.~5 in~\cite{Zweibel2016}.
Here, Figure~\ref{fig-nanoflares-reconnection} describes a unifying
pattern that attempts to explain both the diffuse component of the corona and
the clearly distinguishable coronal loops. Over most of the corona, low-energy
nanoflares occur at a medium to high frequency, and temperature fluctuations
are insignificant. Footpoint driving replenishes the magnetic energy extracted
by each modest event in near real time, producing a statistical steady state.
This is a diffuse corona: in the upper part, the delay between successive
nanoflares is much less than the cooling time, so the temperature fluctuates
around the average. This delay is called high frequency heating.

When a magnetic loop is formed in the diffuse corona, this means that the
reconnection events increase sharply, forming high-energy nanoflares, while the
temperature rises sharply. In this case, the delay between nanoflares is much
longer than the cooling time, and the filament is completely cooled before
reheating. It is the low-frequency heating that creates distinct loops.

Hence, the overall picture of magnetic reconnection is as follows. If the
plasma attracts magnetic loops, which gradually come closer to each other,
causing magnetic reconnection, then nanoflares appear with one of two types of
event frequencies.

From here arises the question of what physics is behind the the two
different event frequencies (see Fig.~\ref{fig-nanoflares-reconnection}).
In the first case, when the plasma interacts with photons of axion origin, 
magnetic reconnection of the nanoflare is the result of low-frequency heating.
As a result, rather rare high-energy nanoflares occur with a low frequency, and
the temperature rises sharply to about 10~MK. On the other hand, when the plasma
(in magnetic reconnection) exists without the influence of photons of axion
origin (see Fig.~\ref{fig-nanoflares-reconnection}), there are frequent
low-energy nanoflares, and the temperature fluctuations are insignificant.

Now let us get back to the second component in Eq.~(\ref{eq06-33}). The
spectrum of solar photons below 10~keV for the active and quiet Sun
(Fig.~\ref{fig06}a), reconstructed from accumulated ROSAT/PSPC observations,
can be described by the second (harder) component ($2~keV < E < 10~keV$), which
is degraded to Compton scattering for a column density above the initial
conversion site of 16$g/cm^2$ (see Fig.~10 in~\cite{Zioutas2009}), while the
initial energy distribution was the converted solar energy spectrum of axions
(Fig.~\ref{fig06}b,c).

In this regard we suppose that the application of the power-law distributions 
with semi-heavy tails leads to a soft attenuation of the observed corona 
spectra (which are not visible above $E > 2 ~keV$), and thus to a close
coincidence between the observed solar spectra and $\gamma$-spectra of axion 
origin (Fig.~\ref{fig06}a,b,c), i.e.

\begin{equation}
\left( \frac{d \Phi}{dE} \right)^{(*)} \approx
\left( \frac{d \Phi _{\gamma}}{dE} \right)^{(*)}_{axions} =
\left( \frac{d \Phi _{\gamma}}{dE} \right)_{CAS} + 
\left( \frac{d \Phi _{\gamma}}{dE} \right)_{compton} ,
\label{eq06-35a}
\end{equation}

\noindent
where the first component on the right is the converted solar axion spectrum
(CAS), and the second one is the degraded spectrum due to multiple Compton
scattering. The spectrum due to inverse Compton scattering (see 
Fig.~\ref{fig06}b,c) is neglected here.

It means that the physics of the formation and ejection of the $\gamma$-quanta 
above $2 ~keV$ through the sunspots into corona is not related to the 
magnetic reconnection theory by e.g. \cite{Shibata2011}, and on the one
hand, it may produce the converted solar axion spectrum, i.e. the differential
$\gamma$-spectrum of axion origin:

\begin{align}
\left( \frac{d \Phi _{\gamma}}{dE} \right) _{CAS} &= P_{a \rightarrow \gamma}
\frac{d \Phi _{a}}{dE} ~~ cm^{-2} s^{-1} keV^{-1} \approx \nonumber \\
& \approx 6.1 \cdot 10^{-3} P_{a \rightarrow \gamma} \frac{d \Phi _{a}}{dE}
~ ph\cdot cm^{-2} s^{-1} bin^{-1} ,
\label{eq3.34}
\end{align}

\noindent
where the spectral bin width is 6.1~eV, $P_{a \rightarrow \gamma}$ is the
probability describing the relative portion of photons of axion origin
channeling along the magnetic tubes, according to~\cite{Peres2000},

\begin{align}
\frac{d \Phi _a}{dE} &= 6.02 \cdot 10^{10} 
\left( \frac{g_{a\gamma}}{10^{10} GeV^{-1}} \right)^2 E^{2.481} \times 
\nonumber \\
&\times \exp \left( - \frac{E}{1.205} \right) ~~cm^{-2}
s^{-1} keV^{-1} ,
\label{eq3.33}
\end{align}

\noindent
is the part of the differential solar axion flux at the 
Earth~\citep{Andriamonje2007}.

On the other hand, a single degraded spectrum due to multiple Compton
scattering is also shown for the column density above the initial
transformation point of $16~g/cm^2$, which actually agrees with the
observed shape of the spectrum (see Fig.~\ref{fig06}a,b,c).

This leads to one important assumption. Since the frequently occurring
low-energy nanoflares are practically not associated with photons of axion
origin (see Fig.~\ref{fig06}b,c and 
Fig.~\ref{fig-corona-compton}c), we assume that the nanoflare
temperature is of the order of $<10^6~K$ (see 
Fig.~\ref{fig-nanoflares-reconnection}).

This means that, neglecting the effect of nanoflares, from the point of
view of the axion mechanism, we should consider the spectra during active and
quiet phases, in which the spectrum is dominated by photons of axion origin,
ejected from magnetic tubes, from the photosphere to the corona (see 
Fig.~\ref{fig06}a,b,c).

\subsection{Scenario of coronal heating by means of axion origin photons}
\label{sec-corona-heating-scenario}

The main task here is to the study of the effect of virtually empty magnetic
tubes in the convective zone and dark matter phenomena of solar axions.

Ultimately, the most important stages include not only the processes of
magnetic energy release, in which magnetic energy is converted into heat,
energetic particles and radiation, but also the plasma reaction to heating,
in which it is organized using degraded spectra due to multiple Compton
scattering closely related to the density columns above the initial conversion
site (Fig.~\ref{fig06}b,c), and the initial energy distribution, that is,
the converted solar axion spectrum (Fig.~\ref{fig06}b,c).

Let us remind here that in our view, the main effective mechanism for heating the solar corona is
the emission of axions from the solar core with an energy spectrum with the maximum
of about 3~keV and the average of 4.2~keV. These axions are supposed to convert
into soft X-rays in very strong transverse magnetic field of an almost empty
tube at the base of the convective zone 
(\mbox{\ref{appendix-b}}). Eventually we suppose that the X-rays,
through the axion-photon conversions in the magnetic O-loop near the
tachocline, channel along the ``cool'' region of the Parker-Biermann magnetic
tube (Fig.~\mbox{\ref{fig-lampochka}}) and effectively supply the necessary
photons of axion origin ``channeling'' in an almost empty magnetic tube to the
photosphere while the convective heat transfer is heavily suppressed 
(\mbox{\ref{appendix-b}}).

The strong magnetic field (with $B \sim 10^7 ~G$) results not only in the
strong suppression of convective heat transfer, but also in the conversion (in
a magnetic O-shaped loop) of high-energy photons (from the radiation zone) into
axions in the tachocline. The appearance of axions in the convective zone leads
(then, and only then!) to the suppression of convective heat transfer. Thus the
mean free path of a photon of axion origin in the convective zone is the result
of the average length from the tachocline to the photosphere. The complete
physics of this process is described in \mbox{\ref{appendix-b}}.

\begin{figure}
\includegraphics[width=\linewidth]{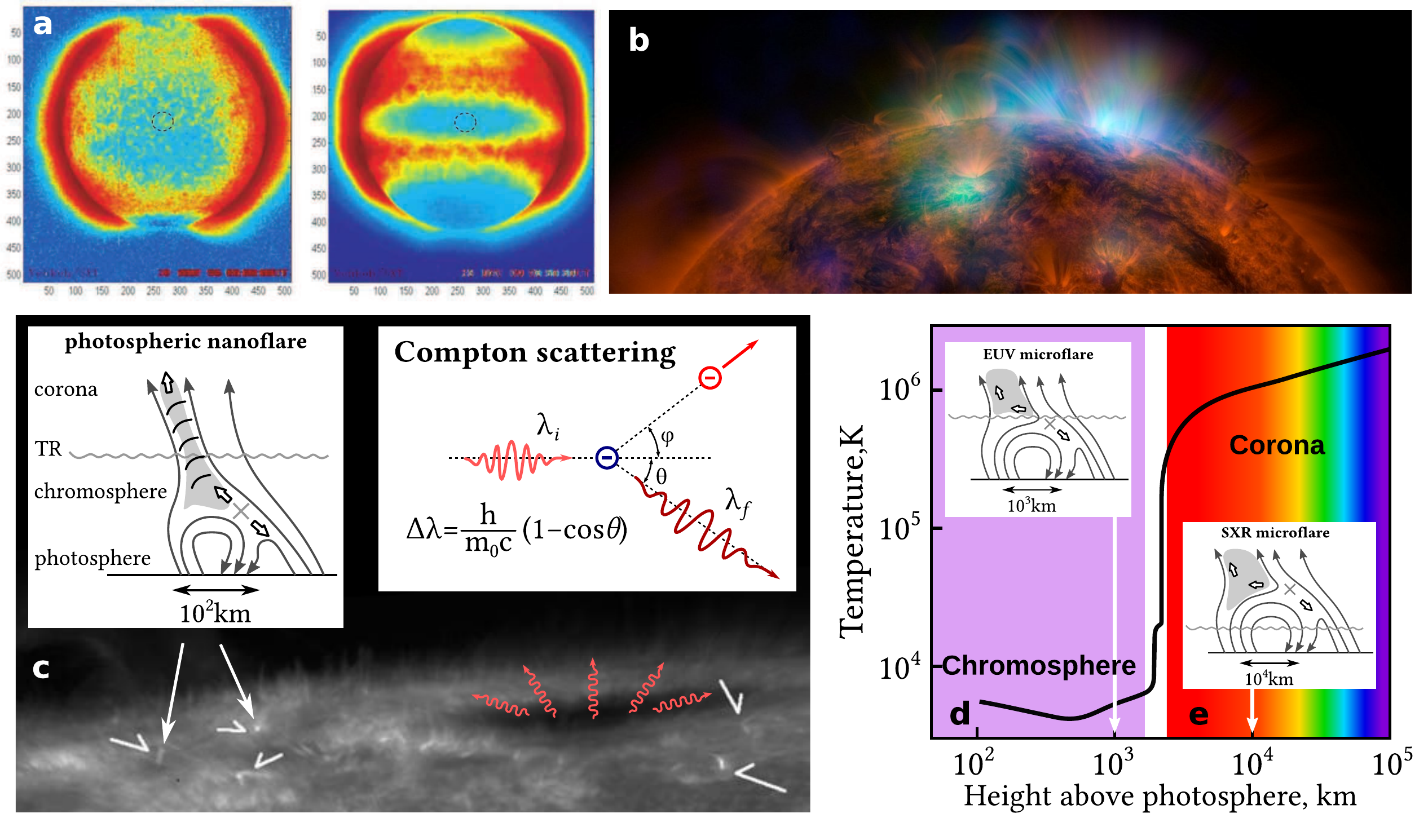}
\caption{\textbf{(a)} Solar images at photon energies from 250~eV up to a few
keV from the Japanese x-ray telescope Yohkoh (1991–2001) (see e.g. Fig.13 
in~\cite{Zioutas2009}). On the left, a composite of 49 of the quietest solar
periods during the solar minimum in 1996. On the right, solar X-ray activity
during the following maximum of the 11-year solar cycle.
\textbf{(b)} NuSTAR data, seen in green and blue, reveal solar high-energy
emission (green shows energies between 2 and 3~keV, and blue shows energies
between 3 and 5~keV). The high-energy X-rays come from gas heated to above 3
million degrees. The red channel represents UV light captured by SDO at
wavelengths of 171\AA, and shows the presence of lower-temperature material in
the solar atmosphere at 1 million degrees (NASA/JPL-Caltech/GSFC, 2014).
\textbf{(c)} Ca II H broadband filter snapshot image of the active region on 17
December 2006 near the west limb, taken with Hinode/SOT. The dark elongated
ellipse seen near the limb is a sunspot (with umbra and penumbra). Numerous
tiny jets ejected from bright points (arrows) can be seen. The footpoints of
these jets are not a simple bright point, but show a cusp- or inverted Y-shape.
This is the characteristic shape of X-ray jets, known as anemone jets (derived
from Fig.~1 in~\cite{Shibata2007}). Schematic illustration of axion origin 
photons flux from the umbra of sunspot (adapted from Fig.~1 
in~\cite{Shibata2007}), and the Compton scattering formula
explaining the formation of a high electron temperature in the corona.
\textbf{(d)},\textbf{(e)} Atmospheric temperature distribution near the solar
surface (derived from Fig.~12 in~\cite{Zioutas2009}).
Insets in (c),(d),(e) with transition regions (TR) illustrate the magnetic
reconnections that occur at different altitudes, from the heights of the
photosphere (c), continuing the chromosphere (d), and ending with the corona
(e) (derived from Fig.~3 in~\cite{Shibata2007}).
}
\label{fig-nanoflares-NuSTAR}
\end{figure}

Since we understand that a practically empty magnetic tube in the photosphere
becomes a sunspot, then a question arises about the observational data from
such magnetic tubes in the solar atmosphere (i.e. from the photosphere to the
corona), from which photons are emitted. Curiously enough, the NuSTAR data 
highlighted in green (2 to 3~keV) and blue (3 to 5~keV) colors) reveal the high
energy solar X-rays (see \mbox{Fig.~\ref{fig-nanoflares-NuSTAR}b}),
which clearly suggest the emission of photons of axion origin from the sunspots.

From here we understand that nanoflares and photons of axion origin in the
corona can be born as nanoflares in penumbra (see 
Fig.~\ref{fig-nanoflares-NuSTAR}c,d,e) and photons of axion origin in the
umbra of the photosphere (see Fig.~\ref{fig-nanoflares-NuSTAR}c).

We first ``observe'' photons of axion origin in the umbra of the photosphere
(see Fig.~\ref{fig-nanoflares-NuSTAR}c) based on images of the Sun at photon
energies from 250~eV to several keV, obtained from the Japanese X-ray telescope
Yohkoh (1991-2001) (see Fig.~\ref{fig-nanoflares-NuSTAR}a), which show X-ray
activity during the last minimum (in 1996) and maximum of the 11-year solar
cycle. On the other hand, the first obtained image of the Sun by NASA's Nuclear
Spectroscopic Telescope Array (NuSTAR), made in high-energy X-rays, provides
answers to questions about the extremely high temperatures observed over
sunspots -- cold dark spots in the Sun (see Fig.~\ref{fig-nanoflares-NuSTAR}b).

When an almost empty magnetic tube rises to the photosphere, the photons of
axion origin, passing the upper boundary of the photosphere, are weakly
scattered in the chromosphere, and then, passing the chromosphere, they are
strongly scattered in the corona (Fig.~\ref{fig-nanoflares-NuSTAR}d,e). 

Recall that the essence of Compton scattering is the interaction of photons of
axion origin with electrons of the outer layers of the Sun. Subsequently, this
leads to an increase in the electron temperature and, as a consequence, to the
heating of the plasma, the temperature of which is spatially equalized in the
corona, and not only in the coronal loops. Moreover, in contrast to the local
dissipation of nanoflares (Fig.~\ref{fig-nanoflares-NuSTAR}c), the photons of
axion origin, the ``observed'' energies of which are associated with red
(0.3-1.0~keV), green (1.0-3.0~keV) and blue (3.0-10~keV) colors from inner
spots (Fig.~\ref{fig-nanoflares-NuSTAR}b), have a global space-time
distribution (0.3-1.0~keV from Fig.~\ref{fig-nanoflares-NuSTAR}b). This is
directly related to the heating rate by photons of axion origin. In this case,
the corona corresponds to an average temperature of about 1~MK with a red
(0.3-1.0~keV) color (Fig.~\ref{fig-nanoflares-NuSTAR}b). The nonlocal
distribution of ``observed'' energies with green (1.0-3.0~keV) and blue
(3.0-10~keV) colors (Fig.~\ref{fig-nanoflares-NuSTAR}b) corresponds to the
temperature above 3~MK. The global distribution does not contradict the fact
that the corona is still hot, like 1~MK, even if there are no spots at some
places (Fig.~\ref{fig-nanoflares-NuSTAR}b).

Next, we are interested in the energy of photons of axion origin in the
chromosphere and corona. The simplified answer is as follows.
We know that the density of electrons in the chromosphere is higher than in the
corona. When electrons in the chromosphere interact with photons of axion
origin, the inverse Compton scattering increases the energy of the photons
(to the right of the average axion energy in Fig.~\mbox{\ref{fig-corona-compton}a})
and thus decreases the energy of the electrons.
A decrease in the energy of electrons
leads to a decrease in temperature, and therefore, in the speed of electrons in
the chromosphere (Fig.~\mbox{\ref{fig-nanoflares-NuSTAR}d}). Conversely, when coronal
electrons interact with photons of
axion origin, the degraded Compton scattering decreases the energy of photons
(to the left of the average axion energy in Fig.~\mbox{\ref{fig-corona-compton}a})
and thus increases the energy of the electrons. Therefore, we understand that
an increase in the energy of electrons leads to an increase in temperature and,
consequently, the speed of electrons in the corona 
(Fig.~\mbox{\ref{fig-nanoflares-NuSTAR}e}). As a result, the
accelerated high-speed electrons and protons, and generally, the continuous
plasma flow propagating almost radially from the Sun and filling the Solar
system up to the heliocentric distances of $\sim~100~a.u.$ (solar wind) is
formed through the gas-dynamic expansion of the corona to the interplanetary
space. At high temperatures in the solar corona ($1.5-4\cdot 10^6~K$) the
pressure of overlying layers cannot compensate the gas pressure of the corona
matter, and the corona expands. In such case the runaway solar wind moves along
the open lines of magnetic field passing through the so-called coronal hole.
The coronal holes are the areas in solar corona which are much less dense and
less hot than the most of the corona, and therefore they are very thin. This
promotes the solar wind, since it is easier for the particles to break through
the chromosphere.

Finally, in \mbox{Section~\ref{sec-corona-heating-solution}} we show that
the number of sunspots is equivalent to the number of magnetic flux tubes, and
thus correlates with the fluxes of axion-origin photons, which in their turn
correlate with the corona heating.

\subsection{Solving the problem of corona temperature}
\label{sec-corona-heating-solution}

Below we show how the plasma reaction to
heating is organized through the interactions of the photon flux of axion
origin with electrons in the outer layers of the Sun, at which a temperature
above $10^6~K$ appears in the corona (see Eqs.~(\ref{eq7.5-sep-18})
 - (\ref{eq7.5-sep-19}); Fig.~\ref{fig06}b,c,d)!

We assume that
the probability $P_{\gamma}$ which describes the relative fraction
of photons of axion origin propagating along the magnetic tube, and 
consequently, describes the relative fraction of the spot area, can be
determined e.g. using $(P_{\gamma})_{max}$ at the maximum solar luminosity:

\begin{equation}
(P_{\gamma})_{max} = \left( P_{a \rightarrow \gamma} \right)_{max} \cdot \dfrac{2 \left \langle sunspot ~area \right \rangle _{max}}
{(4/3) \pi R_{Sun}^2} \sim 0.74 \cdot 10^{-2},
\label{eq7.5-sep-11}
\end{equation}

\noindent and during the minimum of Sun luminosity as:

\begin{equation}
(P_{\gamma})_{min} = \left( P_{a \rightarrow \gamma} \right)_{min} \cdot \dfrac{2 \left \langle sunspot ~area \right \rangle _{min}}
{(4/3) \pi R_{Sun}^2} \sim 8.6 \cdot 10^{-4},
\label{eq7.5-sep-12}
\end{equation}

\noindent where $\langle sunspot ~area \rangle _{max} \approx 7.5 \cdot 10^9 ~km^2 \approx 2470 ~ppm$, $\langle sunspot ~area \rangle _{min} \approx 9.0 \cdot 10^8 ~km^2 \approx 300 ~ppm$ of visible hemisphere is the sunspot area (over the visible hemisphere~\citep{Dikpati2008,Gough2010}) for the cycle 22 experimentally observed by the Japanese X-ray telescope Yohkoh (1991) (see \cite{Zioutas2009}); $R_{Sun} = 6.96 \times 10^5 ~km$; $P_{a \rightarrow \gamma} \approx 1$ at maximum of solar luminosity, and $P_{a \rightarrow \gamma} \approx 0.95$ at minimum 
(see Eqs.~(119) and (120) in \cite{RusovArxiv2019}).

The product of the total fraction of axions originating from the Sun core and the fraction of the sunspot area (see \cite{Dikpati2008,Gough2010}) yields the total fraction of the corona luminosity, or more precisely, the fraction of photon luminosity of axion origin in the corona.

\begin{equation}
\frac{L_{corona}^X}{L_{Sun}} = \frac{L_a^{*}}{L_{Sun}} \cdot P_{\gamma} ,
\label{eq7.5-sep-13}
\end{equation}

\noindent where $L_{corona}^X / L_{Sun}$ is the fraction of the corona luminosity in the total Sun luminosity,
$L_a^{*} / L_{Sun}$ is the fraction of the axion ``luminosity'', $L_{Sun} = 3.8418 \cdot 10^{33} ~erg/s$ is the solar luminosity~\citep{Bahcall2004}.

The axions luminosity fraction $L_a^{*}$ at the maximum and minimum of the Sun luminosity

\begin{equation}
\frac{(L_a^{*})_{max}}{L_{Sun}} = \frac{L_a - (L_{ADM})_{-}}{L_{Sun}} \sim 3.3 \cdot 10^{-4},
\label{eq7.5-sep-14}
\end{equation}

\begin{equation}
\frac{(L_a^{*})_{min}}{L_{Sun}} = \frac{L_a - (L_{ADM})_{+}}{L_{Sun}} \sim 3.6 \cdot 10^{-5},
\label{eq7.5-sep-15}
\end{equation}

\noindent
are determined by the anticorrelated 11-year ADM density variations (with 
$(L_{ADM})_{+} / L_{Sun} \sim 3.24 \cdot 10^{-4}$ and 
$(L_{ADM})_{-} / L_{Sun} \sim 3.24 \cdot 10^{-5}$)\footnote{Here is an
interesting remark by Rosemary~F.G.~Wyse in her paper ``The dark matter
distribution in the Milky Way galaxy'' (1994): ``\textit{...In any case, one
should bear in mind that just as light may not thace mass on cosmological
scales, local maxima in the stellar luminosity function donor require
corresponding peaks in the stellar mass function, since the mass-luminosity
relationship depends of the details of stellar structure, such as opacity
sources, which are mass-dependent. The solar neighborhood luminosity function
has peaks at $\sim 10^{-3} L_{Sun}$, corresponding to $\sim 0.3 M_{Sun}$, and
at $\sim 0.15 L_{Sun}$, corresponding to $\sim 0.7M_{Sun}$, but a monotonic
initial mass function still provides a good fit, provided one uses the correct
mass-luminosity relation~\cite{Kroupa1990}. Farther, 
Kroupa~et~al.~\cite{Kroupa1990,Kroupa1993} demonstrated that all the models on
the solar neighborhood initial mass function that are consistent with the both
the observed luminosity function and mass-luminosity relation  function
converge when extrapolated to low masses.''}},
which are gravitationally
captured in the solar interior (see Sect.~\ref{sec-dark-matter}).
Asymmetric dark matter provides a natural explanation of the comparable
densities of baryonic matter and dark matter.
The contribution of dark matter to the total mass of the star is totally
negligible, but the presence of dark matter in the stellar interior changes
the local properties of the plasma, and by doing so will affect the evolution
of the star. We present models of dark matter with possible momentum- or 
velocity-dependent interactions with nuclei, based on direct detection and
solar physics. An important note is that this, as a rule, can lead to the
absence of self-annihilation today\footnote{It can be shown
(see~\cite{Pearce2013}) that, if dark matter is both asymmetric and
self-interacting, then its detection in gamma rays is possible. Although
asymmetric dark matter particles do not annihilate, their interactions can
result in emission of quanta of the field that mediates the self-interaction.
As a result, the unique indirect detection signals created by the minimal model
of self-acting asymmetric scalar dark matter are discussed here. Through
the formation of dark matter bound states, a dark force mediator particle may
be emitted; the decay of this particle may produce an observable gamma-signal.
\\
\indent
Thus, non-minimal physics, such as light vector or scalar force mediator in the
dark sector (see analogue of~\cite{Baldes2017,Baldes2018}), can be explained by
the well-known observation of a significant change in the gamma radiation flux
of the solar disk over time during (of magnitude in 1-10~GeV), which appears to
be anticorrelated with solar activity. This is the first clear observation of
such a time variation~\cite{Ng2016}. Nonetheless, the
anticorrelation with solar activity indicates that the bulk of the solar-disk
gamma rays can be explained by cosmic-ray interactions in the solar atmosphere
and the gamma-ray production process is strongly affected by the solar magnetic
fields.},
which allows large amounts of ADM to accumulate in stars like the
Sun~\citep{Vincent2015a}. 
Once the direct detection limits are accounted for, however, the best solution
is spin-dependent $v^2$ scattering with a reference cross-section of
$10^{-35}~cm^2$ (at a reference velocity of $v_0 = 220~km/s$, and a dark matter
mass of about 5~GeV) (see \cite{Vincent2016}).

The ADM particles absorb energy in the hottest,
central part of the core, they then travel to a cooler, more peripheral, area
before the scattering again and redistribute their
energy~\citep{GouldRaffelt1990a}.
This reduces the contrast of temperature across the core region and reduces the
central temperature. A colder core produces less neutrinos from the
temperature-sensitive fusion reactions, and also less axions from the
temperature-sensitive reactions of electron-nucleus collisions (see 
e.g.~\cite{Raffelt1986}). So the fluxes of $^7 Be$ and $^8 B$ neutrino, and
axions (see (\ref{eq7.5-sep-15})) can be
significantly reduced. This is accompanied by a slight change in the $pp$ and
$pep$ fluxes, as required by the constancy of the solar luminosity.

Taking into account the known observed variability of the corona luminosity in the X-ray range recorded in the ROSAT/SPC experiments,

\begin{equation}
\frac{(L_{corona}^X)_{max}}{L_{Sun}} \sim 1.2 \cdot 10^{-6}, ~~~
\frac{(L_{corona}^X)_{min}}{L_{Sun}} \sim 7.0 \cdot 10^{-8},
\label{eq7.5-sep-17}
\end{equation}

\noindent
we can compare the ``experimental'' (on the left; also (\ref{eq7.5-sep-17})) and theoretical equations (on the right; also Eqs.~(\ref{eq7.5-sep-11}) - (\ref{eq7.5-sep-12}) and (\ref{eq7.5-sep-14})-(\ref{eq7.5-sep-15}))

\begin{equation}
\frac{(L_{corona}^X)_{max}}{L_{Sun}} = \frac{(L_a^{*})_{max}}{L_{Sun}} \times
(P_{\gamma})_{max} \sim 2.4 \cdot 10^{-6},
\label{eq7.5-sep-18}
\end{equation}

\begin{equation}
\frac{(L_{corona}^X)_{min}}{L_{Sun}} = \frac{(L_a^{*})_{min}}{L_{Sun}} \times
(P_{\gamma})_{min} \sim 3.1 \cdot 10^{-8},
\label{eq7.5-sep-19}
\end{equation}

\noindent which are good enough, but our estimates of the theoretical
variations of solar luminosity 
$(L_{corona}^X)_{max} \sim 1.0 \cdot 10^{28}~erg/s$ at a
temperature $T_{max} \sim 4\cdot 10^6~K$ and 
$(L_{corona}^X)_{min} \sim 1.2 \cdot 10^{26}~erg/s$ at a temperature
$T_{min} \sim 1.5\cdot 10^6~K$ are still somewhat better than the data in
\cite{Peres2000} (see Fig.~\ref{fig06}d).

At the same time, we must keep in mind that the ``experimental'' data (on the left: Eq.~(\ref{eq7.5-sep-17})) come from the spectra of the corona, synthesized with the MEKAL spectral code directly from the emission measure $EM (T)$ distribution derived with spectral fitting of ROSAT/PSPC spectra, as reported by \cite{Peres2000}
This means that the values of the ``experimental'' data (equation
(\ref{eq7.5-sep-17})) are rather rough due to their uncertainty, while the
theoretical data (Eqs.~(\ref{eq7.5-sep-18})-(\ref{eq7.5-sep-19})) are more
accurate because of the temperature values (see Fig.~\ref{fig06}d), which
depend only on physics of the free energy accumulated by photons of axion
origin in a magnetic field by means of degraded spectra due to multiple
Compton scattering (see Fig.~\ref{fig06}b,c), and which is rapidly released
and converted into heat and plasma motion with a temperature of
$\sim 4 \cdot 10^6 ~K$ at maximum and $\sim 1.5 \cdot 10^6 ~K$ at minimum of
solar luminosity (see Fig.~\ref{fig06}d).

From here we understand that the end result of the processes of the coronal
magnetic energy release, which is converted into heat, energetic particles and
radiation, would be described by a temperature of less than $10^6~K$ (see 
Eqs.~(\ref{eq06-34})-(\ref{eq06-35}); Fig.~\ref{fig06}b,c).
Meanwhile the
reaction of the plasma to heating, in which it is realized by the interactions
of the flux of the axion-originated photons with electrons in outer layers of
the Sun, would be described by a temperature above $10^6~K$ (see 
Eqs.(\ref{eq7.5-sep-18})-(\ref{eq7.5-sep-19}); Fig.~\ref{fig06}b,c,d)!

Let us note another interesting point. In contrast to the variability of corona
luminosity (on the left Eqs.~(\ref{eq7.5-sep-20a}) and (\ref{eq7.5-sep-20b})),
according to the ``experimental'' ASCA/SIS data,

\begin{equation}
\frac{(L_{corona}^X)_{max}}{L_{Sun}} \sim 0.39 \cdot 10^{-6} \neq \frac{(L_a^{*})_{max}}{L_{Sun}} \cdot (P_{\gamma})_{max} \sim 2.4 \cdot 10^{-6} ,
\label{eq7.5-sep-20a}
\end{equation}

\begin{equation}
\frac{(L_{corona}^X)_{min}}{L_{Sun}} \sim 0.6 \cdot 10^{-8} \neq \frac{(L_a^{*})_{min}}{L_{Sun}} \cdot (P_{\gamma})_{min} \sim 3.1 \cdot 10^{-8},
\label{eq7.5-sep-20b}
\end{equation}

\noindent
the strongest results is the theoretical variability of corona luminosity (on
the right: Eqs.~(\ref{eq7.5-sep-20a})-(\ref{eq7.5-sep-20b})), which coincide with
the variations of the corona luminosity according to ROSAT/PSPC quite well (on
the left: Eqs.~(\ref{eq7.5-sep-18})-(\ref{eq7.5-sep-19})).

The preliminary result is as follows. If the hadron axions, arising from the
Sun core, are controlled by 11-year variations of ADM density, that is
gravitationally captured in the Sun interior (see 
Sect.~\ref{sec-dark-matter}), then the hard part of the spectrum of solar
photons has the axion origin, which is determined mainly by theoretical
estimate as the product of the fraction of axions in the Sun core and the
fraction of the sunspot area (see \cite{Dikpati2008,Gough2010}). The latter is
the result of the Sun luminosity variations (see 
Eqs.~(\ref{eq7.5-sep-18})-(\ref{eq7.5-sep-19})). This means that the Sun
luminosity variations and other solar cycles are the result of 11-year ADM
density variations, which are anticorrelated with the density of hadron axions
in the core of the Sun.

So on the basis of photons of axion origin, we obtain two significant results.
First, the solar luminosity of the corona is absolutely identical to the
luminosity of the photosphere (which is confirmed by 
Eqs.~(\ref{eq7.5-sep-18})-(\ref{eq7.5-sep-19})), since the number of photons of
axion origin, which exit through the magnetic flux tubes of the photosphere,
despite the change in their density, is almost equal to the number of photons
in the low-density ambient plasma of the corona.
As a result, the experimental ratio $L_{corona}^X / L_{Sun}$ 
(see Fig.~\ref{fig-corona-compton}a)
roughly coincides with the theoretical ratio $L_a^{*} / L_{Sun} \times P_{\gamma}$
(see e.g. Eqs.~(\ref{eq7.5-sep-18})-(\ref{eq7.5-sep-19})) in the minimum and
maximum luminosity of the corona.

Second, the energy distribution of the emitted axions is not like a blackbody
spectrum, since the spectrum of the incident photons of axion origin is
modulated by the frequency dependence of the cross section. As we have already
shown, for a typical solar spectrum, the maximum of the axion differential flux
occurs at $E_a / T \approx 3.5$, while the average axion energy 
$\langle E_a / T \rangle \approx 4.4$ \citep{Raffelt1986}. This means
that the energy of the average photon of axion origin can generate a
temperature of the order of $T_a \sim 1.11 \cdot 10^7 ~K$ (see e.g. 
Fig.~\ref{fig-corona-compton}a,c) under certain conditions of coronal
substances, which is close to the temperature $T_{core} \sim 1.55 \cdot 10^7~K$
of the solar core!
As a result, the free energy accumulated by the photons of axion origin in a
magnetic field by means of degraded spectra due to multiple Compton scattering
(see Fig.~\ref{fig06}b,c), is quickly released and converted into heat and
plasma motion with a temperature of $\sim 4 \cdot 10^6 ~K$ at maximum and
$\sim 1.5 \cdot 10^6 ~K$ at minimum of solar luminosity
(see Fig.~\ref{fig06}d). 
Under some conditions, the photons of axion origin with very
powerful flares in the corona can partially generate energy,
which corresponds to a temperature of the order of 
$T_a \sim 10~MK$ (see e.g. Fig.~\ref{fig-corona-compton}b,d).

As a result, according to our physical understanding, the coincidence of the
theoretical (see e.g. Eqs.~(\ref{eq7.5-sep-18})-(\ref{eq7.5-sep-19})) and
experimental (see ROSAT/PSPC) photon spectra of the corona 
is connected, on the one hand, with the appearance of the magnetic flux and
simultaneously the flux of photons of axion origin in the outer layers of the
Sun, and on the other hand, with the manifestation of the basic mechanism of
formation of sunspots and active regions correlated with the solar cycle, which
determines the formation of the corresponding variation in the energy release
of the axion-originated photons in the solar corona.

Curiously enough, we have understood that the fundamental physics of the Sun,
as if recalling a forgotten, but physically profound question, raised long ago
by R.H.~Dicke, ``Is there a chronometer hidden deep in the 
Sun?''~\citep{Dicke1978,Dicke1979,Dicke1988}, suggests the possibility of the
existence of ADM (as a result of the energy conservation law with the Galactic
frame velocity, density and dispersion) in the Sun core, with which, for
example, variations in the ADM density and, as a consequence, variations in the
neutrino flux, the solar cycle, solar luminosity, sunspots and other solar
activities, seem to be paced by an accurate ``clock'' inside the Sun.

The question is very simple: If the ADM density variation controls the ``clock''
inside the Sun, then what is the physics behind this process?

\section{How the ADM density variation around the black hole controls the ADM density variation inside the Sun}
\label{sec-dark-matter}

Let us try to find an answer to the important, yet unconventional question: is there an observable connection between the 11-year variations of ADM density in the solar interior and the periods of S-stars revolution around the SMBH at the center of the Milky Way (see e.g. \cite{Genzel2010,Dokuchaev2015,Mapelli2016})?

\begin{figure*}[tbp!]
  \begin{center}
    \includegraphics[width=18cm]{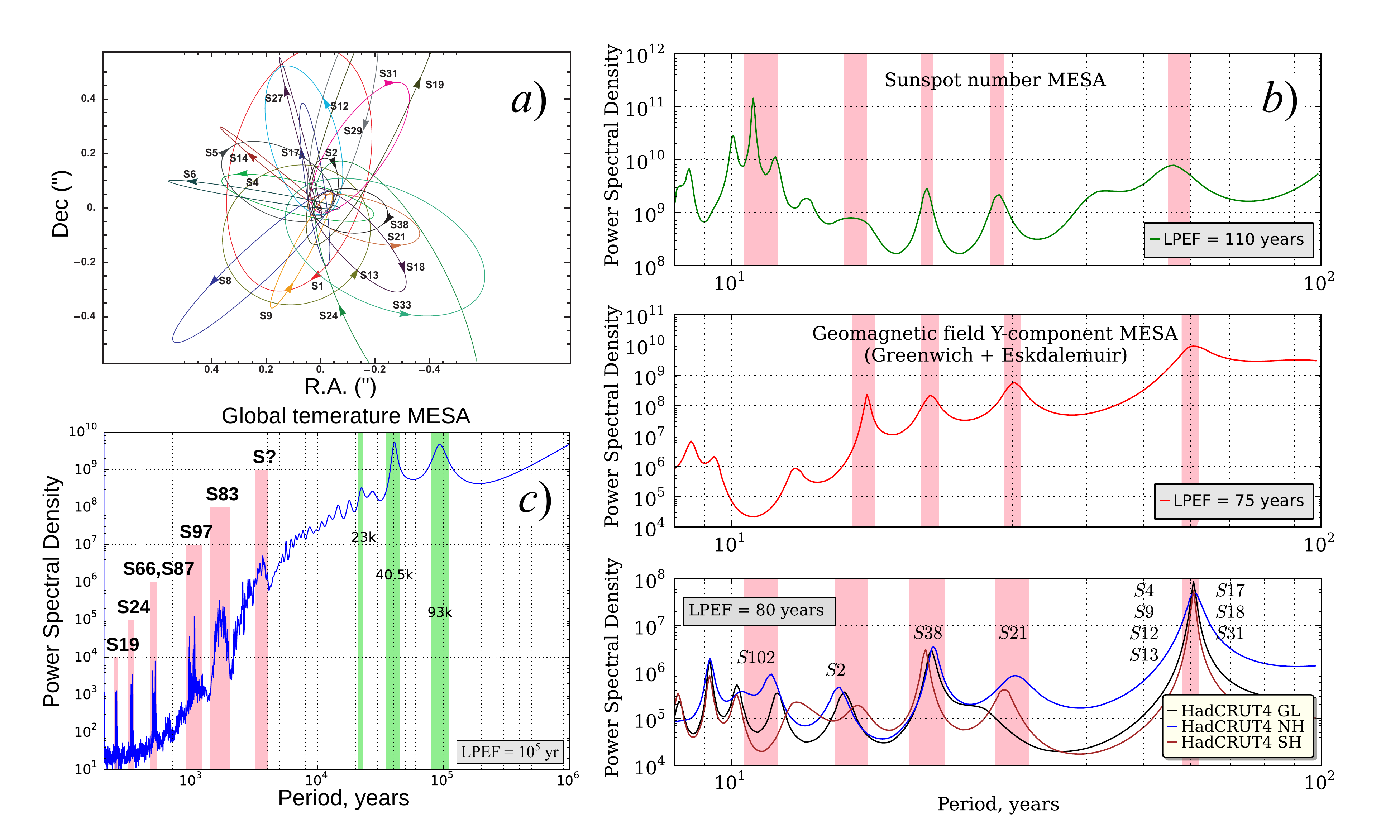}
  \end{center}
\caption{\textbf{(a)} Stellar orbits at the GC in the central arcsecond (declination Dec(”) as a function of time for the stars and red ascension R.A.(”)) \citep{Gillessen2009,Genzel2010}. The coordinate system is chosen so that Sgr A* (the SMBH with the mass $\sim 4.3 \cdot 10^6 ~M_{Sun}$ ) is at rest. 
\textbf{(b)} Power spectra of the sunspots (1874-2010 from Royal Greenwich Observatory), geomagnetic field Y-component (from Greenwich and Eskdalemuir observatories, \cite{WDC2007}) and HadCRUT4 GST (1850-2012) (black), Northern Hemisphere (NH) and Southern Hemisphere (SN) using the maximum entropy spectral analysis (MESA); red boxes represent major astronomical oscillations associated to the major heliospheric harmonics associated to the orbits of the best known short-period S-stars (S0-102, S2, S38, S21, S4-S9-S12-S13-S17-S18-S31) at the GC \citep{Gillessen2009,Genzel2010,Gillessen2013,Gillessen2017} and to the solar cycles (about 11-12, 15-16, 20-22, 29-30, 60-61 years).
\textbf{(c)} Power spectra of the global temperature as reconstructed by \cite{Bintanja2008}; red boxes represent the major astronomical oscillations associated to the major heliospheric harmonics associated to the orbits of the best known non-short-period S-stars (S19, S24, S66-S87, S97, S83, S?) at the GC \citep{Gillessen2009,Genzel2010} and to the solar cycles (about 250, 330, 500, 1050, 1700, 3600 years); green boxes represent the major temperature oscillations presumably associated with the variations of the Earth orbital parameters: eccentricity ($\sim$93~kyr), obliquity ($\sim$41~kyr) and axis precession ($\sim$23~kyr).}
\label{fig-stellar-orbits}
\end{figure*}

On basis of the experimental data (see Fig.~\ref{fig-stellar-orbits}) on the oscillations of the sunspot number, the geomagnetic field Y-component and the global temperature we explicitly demonstrate that their periods coincide with revolution periods of S-stars orbiting a SMBH at the GC of the Milky Way. It is absolutely obvious that such a fine coincidence cannot be random. Then the next quite natural question arises: how do the solar and terrestrial observables ``know'' about motion of S-stars? And a hypothesis inevitably comes to mind: the ``carrier'' is none other than DM. More specifically, S-stars can modulate DM flows in our galaxy and, consequently, cause variations of DM space and velocity distributions, in particular, at the Sun and Earth positions. Further, these variations may cause the corresponding variations of the Solar System observables by means of some mechanism, e.g. the interaction of DM particles, which correlate with baryonic matter, with the cores of the Sun and the Earth. Such a probable mechanism is a subject of the current section. Here our aim is to stress that the available experimental data indicate the frequency transfer from the center of our galaxy to the Solar System. This fact can serve as an indirect evidence of the proposed hypothesis that DM plays the role of the variations carrier.

In order to answer this question, let us first consider all unexpected and intriguing implications of the 11-year modulations of ADM density in the solar interior and ADM around the BH. 

The essence of fundamental magnetic processes associated with quantum gravity and the generalized thermomagnetic EN effect in the “tachocline” near the BH boundary may be described rather simply as follows. According to our understanding, one of the fundamental effects of the holographic principle of quantum gravity is the existence of ``tachoclines'' in all stellar objects in the Universe, including all galaxies and, of course, our Sun, magnetic white dwarfs, neutron stars and BHs of the Milky Way
(see~\ref{appendix-c}).

\begin{figure*}[tbp]
  \begin{center}
     \includegraphics[width=15cm]{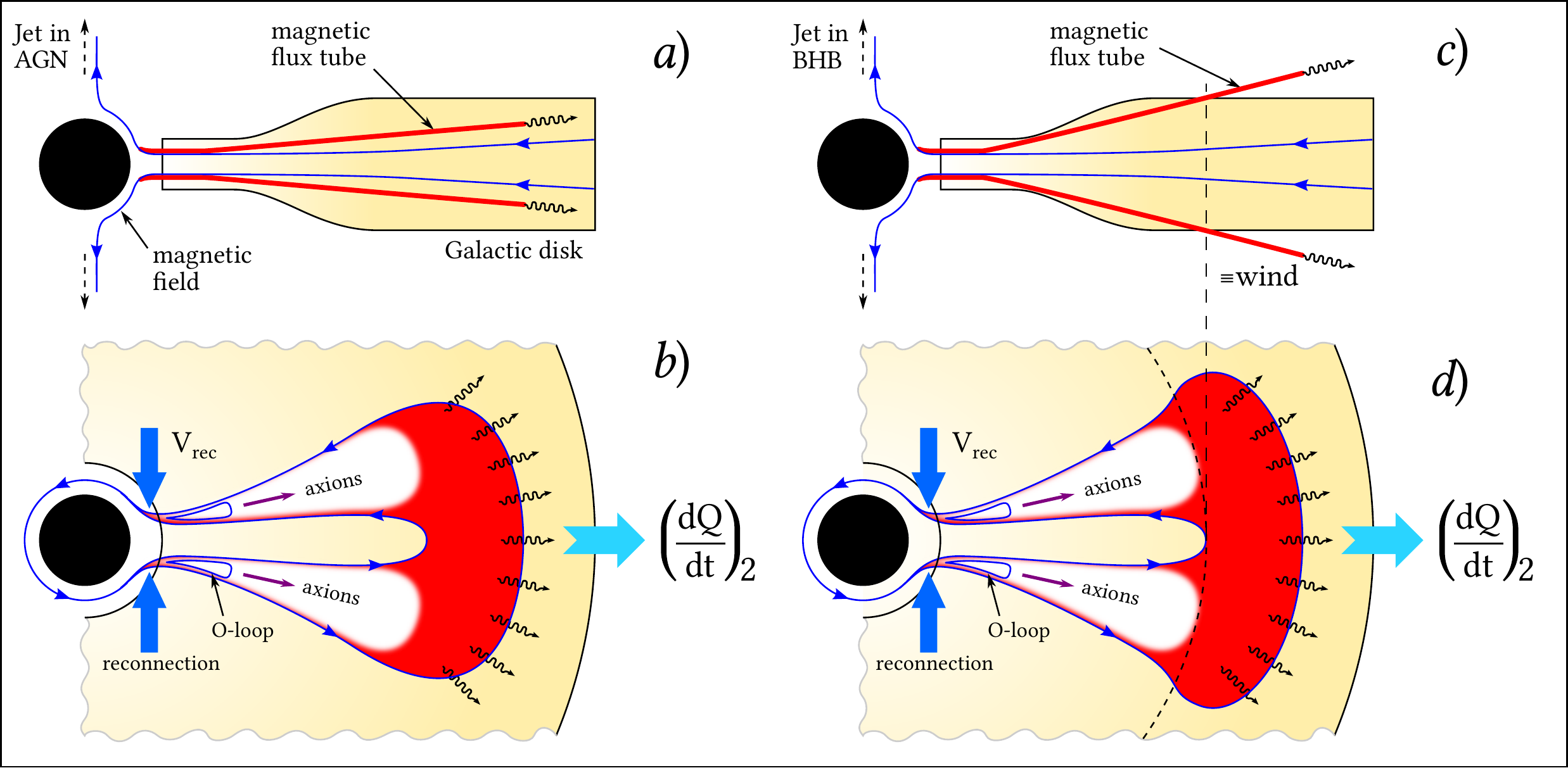}
  \end{center}
\caption{Schematic representation not to scale: Generalized thermomagnetic
EN effect and virtually empty MFT born
anchored to the BH tachocline and risen to the disk surface by the
neutral buoyancy ($\rho_{ext} = \rho_{int}$; see also Figs.~9 and~11 
in~\cite{RusovArxiv2019}). If the virtually empty magnetic flux tubes
(\ref{appendix-b})
 are born
with strong fields \textbf{(a)}, then the flux tubes practically do not reach
outside of the disk, while with less strong fields \textbf{(c)}, the flux tubes
go out of the disk at a small angle. So the visible (c,d) or invisible (a,b)
buoyant flux tubes are the analog of the variation or disappearance of the
wind through the various states of BH X-ray binaries (BHBs), which are
interpreted as a variation in the driving mechanism of the wind
(see \cite{Chakravorty2016} and Refs. therein). Based on the generalized
thermomagnetic EN effect, the buoyancy of MFTs
with the toroidal magnetic field $\gg 10^5 ~G$ in the BH tachocline,
ultimately, through $\nabla \rho$-pumping 
(see Fig.~17,19 in~\cite{RusovArxiv2019})
 with the magnetic field 
$\sim 10^5 ~G$ and the dominant Coriolis force creates an upward curved magnetic
loop with a tilt angle from the Joy's law (see (a), (c); see also Sect.~3.1.3.2
and Fig.~19a in~\cite{RusovArxiv2019}). The slope at low latitudes is near the
disk plane (see (a), (c); see also Fig.~16a,b in~\cite{RusovArxiv2019}). Here
the keV photons ((a)-(d); see analogous Fig.~11a in~\cite{RusovArxiv2019}),
coming into the tachocline from the solar-like radiation zone, are turned (b,d)
into axions by means of the horizontal magnetic field of the O-loop ((b,d); see
analogous Figs.~6a,~9a and~11a in~\cite{RusovArxiv2019}). Some small photon flux
can still pass through the ``ring'' between the O-loop and the tube walls (see
Figs.~6a and~9a in~\cite{RusovArxiv2019}) and reach the solar-like penumbra of
the MFT in the disk (see Fig.~9 and Sect.~3.1.3 in~\cite{RusovArxiv2019}). The
physics and the possible estimate of the radiative heating $(dQ/dt)_2$ (see e.g.
Fig.~11a in~\cite{RusovArxiv2019};
\ref{appendix-b})
passing through the ``ring'' of the magnetic
tube (b,d), the speed of the reconnection $V_{rec}$ (see analogous Fig.~12
in~\cite{RusovArxiv2019};
\ref{appendix-b}),
O-loops (see Figs.~6a, 9a, 10d, 11a 
in~\cite{RusovArxiv2019}), and the lifetime of the magnetic tube are presented
in Sect.~3.1.3.1 in~\cite{RusovArxiv2019}.
}
\label{fig-GalacticDisk-MagTube}
\end{figure*}

Let us recall that the solar radiation zone rotates approximately like a solid,
and the convection zone has a differential rotation. This leads to the
formation of a very strong shear layer between these two zones, called the
tachocline. Similar physics of the tachocline exists in a BH. As a
result, the tachocline shear layers produce virtually empty MFTs
(see~\ref{appendix-b} and Sect.~3.1.2 in \cite{RusovArxiv2019}), anchored in the BH tachocline
and rising to the surface of the disk (see 
Fig.~\ref{fig-GalacticDisk-MagTube}b,d). Since it is known that disk parts
rotate around a BH at different speeds that reinforce the fields
(Fig.~\ref{fig-GalacticDisk-MagTube}a), this turns the accretion disk into
a vortex, pulling the substance into a BH and fueling winds that blow
some of it outwards (see Fig.~\ref{fig-GalacticDisk-MagTube}b).

So, if the physics of winds in the disk is identical to the physics of
practically empty MFTs, which start from the tachocline and rise
to the outer (Fig.~\ref{fig-GalacticDisk-MagTube}c,d) or inner
(Fig.~\ref{fig-GalacticDisk-MagTube}a,b) part of the disk, then it means that
the generalized thermomagnetic EN effect produces the magnetic
tubes or the so-called winds and, on the one hand, the substance current
flowing in the direction of the pole of the poloidal (meridional) field of the
tachocline in the form of jets.

If we recall that the variations of the toroidal magnetic field and ADM
density in the BH tachocline exactly anticorrelate with each other
(see Eq.~(\ref{eq07-151})), this means that in strong fields the magnetic tubes
rise only inside the disk, which is equivalent to the disappearance of winds,
and thus they slow down the speed of the accretion. And the opposite, in weak
fields the magnetic tubes rise out of the disk surface (see 
Fig.~\ref{fig-GalacticDisk-MagTube}c), which is equivalent to observing the
disk winds, and thus they increase the speed of the accretion.

When the magnetic pressure is large at low densities and
high temperatures, the low densities of the BH accretion material and
high magnetic fields in the tachocline lead to the formation of highly
collimated spectrally-hard jets, but without a ``visible'' magnetic tube
(jet in AGN in Fig.~\ref{fig-GalacticDisk-MagTube}c). And vice versa, when the 
magnetic pressure is relatively weak at high densities and 
low temperatures, the high density of the BH accretion material and
relatively low magnetic fields in the tachocline lead to the formation of
strongly collimated spectrally-soft jets and a visible magnetic tube, i.e. 
a visible disk ``wind'' (jet in BHB in Fig.~\ref{fig-GalacticDisk-MagTube}c,d).

So, the resulting conclusion looks quite clear: the physical nature of the
generalized thermomagnetic EN effect is the cause of both the
formation of magnetic tubes rising from the BH tachocline to the disk
and the formation of meridional currents in the direction of the BH
pole, which generate jets, for example, in AGN or BHB.

Since the physics of magnetic tubes, rising from the BH ``tachocline''
to the disk, is almost identical, in our view, to the physics of visible or
``invisible'' disk winds, below we consider the common properties of the
generalized thermomagnetic EN effect and the known models of
disk winds. Curiously enough, the latter are associated with the mechanism of
DM variations around a BH.

Such winds may arise from various processes, which makes their sources disputable (see e.g. \cite{Fukumura2017} and Refs. therein).
However, the X-ray spectroscopic data and analysis of the wind associated with the X-ray binary (XRB) GRO J1655-40 \citep{Miller2006a,Miller2006b,Miller2008,Miller2012,Kallman2009,Luketic2010} argued in favour of the magnetic origin, excluding all candidate processes except for the following two: the semi-analytic MHD wind model of \cite{Fukumura2017} and the MHD outflow model of \cite{Chakravorty2016}. Plus our model of the generalized thermomagnetic EN effect.

The observations indicate that disk winds and jets in X-ray binaries are
anticorrelated \citep{Miller2006b,Miller2008,Miller2012,Neilsen2009,King2012a,
King2013,Ponti2012}. This indicates a link between disk properties, magnetic
field configurations and outflow modes, the repulsive magnetic field induced by
the generalized thermomagnetic EN effect (see Eq.~(\ref{eq06-08}), 
(\ref{eq06-14}) in \ref{appendix-a}),
which produces the variations of
the magnetic fields near the BH boundary (see 
e.g.~\cite{Eatough2013,Zamaninasab2014,Johnson2015}),

\begin{align}
\frac{\left[ B_{tacho}^2 (r,t) \right]_{max-cycle}}{\left[ B_{tacho}^2 (r,t) \right]_{min-cycle}} = 
\frac{\left[n_{tacho}^{min} (r,t) \cdot T_{tacho}^{max} (r,t)\right]_{max-cycle}}
{\left[n_{tacho}^{max} (r,t) \cdot T_{tacho}^{min} (r,t)\right]_{min-cycle}} , 
~~~~~ [T (r,t)]^{1/4} n (r,t) = const , 
\end{align}

\noindent
as well as the variations of the ADM density $n_{tacho} ^{ADM}$ and baryon matter $n_{tacho}$,

\begin{equation}
\frac{\left[ n_{tacho}^{min} (r,t) \right]_{max-cycle}}
{\left[ n_{tacho}^{max} (r,t) \right]_{min-cycle}} \approx
\frac{\left[ n_{tacho}^{ADM-min} (r,t) \right]_{max-cycle}}
{\left[ n_{tacho}^{ADM-max} (r,t) \right]_{min-cycle}}\, ,
\label{eq07-150}
\end{equation}

\noindent
determined by the modulations of the ADM density $n_{tacho}^{ADM}$.

It is known (see \cite{Chakravorty2016}) that the X-ray spectra of BH X-ray binaries (BHBs) contain blueshifted absorption lines, which means the presence of outflowing winds. The observations also show that the disk winds are equatorial and they mostly occur in the Softer (disk dominated) states of the outburst, being less in the Harder (power-law dominated) states.

The properties of the accretion disk are used to infer the driving mechanism of the winds (see e.g. \cite{Neilsen2012} and Refs. therein). And more or less prominent winds through the various states of the BHB have been interpreted as a variation in the magnetic driving mechanism of the wind \citep{Miller2006a,Miller2006b,Kallman2009,Neilsen2012}.

In our case the ADM density $n_{tacho}^{ADM}$ modulations, associated with the
generalized thermomagnetic EN effect (see Eqs.~(\ref{eq07-150}),
(\ref{eq07-151})), lead to the variations of the toroidal magnetic fields in
the accretion disk (see Fig.~\ref{fig-GalacticDisk-MagTube}). In order to
understand the main motivation for the toroidal magnetic field variations in
the accretion disk, forming the more or less significant
winds, it is necessary to
discuss the difference between the winds and jets from accretion disks.

Given Eq.~(\ref{eq07-150}) and the high (low) magnetic pressure, the density of
ADM is relatively low (high), and the gas temperature is high (low). This means
that with the high density of ADM and low temperature, the magnetic driving
mechanism produces the accretion disk wind, which is equatorial 
(see~\ref{fig-GalacticDisk-MagTube}b,d) and occurs in the soft states of the BHB
outbursts (see e.g. the analogous model by \cite{Chakravorty2016}).
Alternatively, with the low density of ADM and high temperature, the magnetic
pressure is rather high and, consequently, the magnetic driving mechanism
produces the weak or less prominent wind in the hard states (AGN jets).

Either way, the nonrelativistic disk winds and relativistic jets are anticorrelated, since the relativistic AGN jet, induced by the vertical toroidal magnetic field (see Fig.~\ref{fig-GalacticDisk-MagTube})
and collisions between ADM and nuclei
in the close vicinity of a SMBH (see e.g. the analogous model by \cite{Lacroix2016}), is determined by the modulations of the ADM density (see Eq.~(\ref{eq07-150})):

\begin{equation}
\frac{\left[ B_{tacho} \right]_{max-cycle}}
     {\left[ B_{tacho} \right]_{min-cycle}} =
\left \lbrace 
\frac{\left[ \rho _{tacho}^{ADM-max} (r,t) \right]_{min-cycle}}
     {\left[ \rho _{tacho}^{ADM-min} (r,t) \right]_{max-cycle}}
\right \rbrace ^{3/2} .
\label{eq07-151}
\end{equation}

This eventually means that the strong magnetic fields near the BH boundary are
caused by quantum gravity of the dyonic BH \citep{Hartnoll2007}, which
determines the existence of the generalized thermomagnetic EN effect
(\ref{appendix-a}).
So we understand that the major effect of quantum gravity here is that the
initial acceleration (deceleration) of the disk winds and BHB flares (AGN jets)
originate from less (more) strong magnetic fields in the accretion disk near the
BH boundary (see also Eq.~(\ref{eq07-151}), 
Fig.~\ref{fig-GalacticDisk-MagTube}), which predetermine the
modulations of the ADM density -- the process connected with the variability of
the accretion flows, nonrelativistic disk winds and relativistic BHB or AGN
jets.

This raises the question of how the observational data on AGN (or jet)
variability, which theoretically anticorrelates with a variation of the
toroidal magnetic field in the accretion disk, will be related to the
observational data on variations in the accretion rate or e.g. the magnetic
disk winds? In our opinion, despite the elusiveness of direct observations,
some ideas, for example, of \cite{Rodrigues2015},
about the possible observation of Parker loops from synchrotron radiation of
galaxies near their edges, or from the Faraday rotation, or the idea of
\cite{Chakravorty2016} on the possible observable difference between winds and
jets from accretion disks can provide the indirect observational support.

In this sense, we are very interested in obtaining the indirect observations of
anticorrelation between winds and jets from accretion disks. The point is that
the modulation of a ADM halo leads to the situation when the high density
of ADM corresponds to the disk wind and high accretion rate, while
the low density of ADM corresponds to the low accretion rate and, as a
result, to the jets from the BH. As we understand, in order to get a
generalized picture of the connections between winds and jets from accretion
disks, it is necessary to obtain the indirect observation, which physically
explains the remarkable connection between all mentioned systems: the
modulation of the ADM halo at the center of the galaxy $\rightarrow$
the vertical waves of density from the disk to the solar 
neighborhood~\cite{Purcell2011,Gomez2013,Carlin2013,Widrow2012,Widrow2014,
Xu2015,Gomez2017,Carrillo2018,Carrillo2019,Laporte2018a,Laporte2018b}
$\rightarrow$ variations (cycles) of sunspots $\rightarrow$ the variability of
the local positions of orbital S-stars near the BH.

Our plan is the following. The main goal is to show that the observed
variability of the local positions of the S-stars orbiting near the BH
is an indicator (or dynamic probe) of the disk wind speeds variation or,
equivalently, of the accretion rate variations of the BH.
The intermediate goal is to briefly discuss the formation of the elliptically
orbital distributions and periodic cycles of the S-stars, which are located
in 0.13 ly $\approx$ 0.04 pc (see Fig.~26c in~\cite{RusovArxiv2019}) from the
BH.

Here we present a detailed analysis of the effects of DM capture
and energy transport on the structure and evolution of main-sequence B-stars,
specifically those which might exist at the GC. First we need to
highlight some important modulation properties of the S-stars and ADM around
the BH:

$\bullet$
The greatest
capture happens when the star is farthest from the center of the galaxy, at
apoapsis. This is because it slows down relative to the ADM halo and achieves a
significant capture rate for a time
before turning back towards the BH. By the time it reaches
periapsis, the star is moving so quickly that the capture is essentially zero,
regardless of how high the ADM density is.

$\bullet$ When the S-stars, and especially, for example, S102 and S2, approach
(retreat from) the center of the Milky-Way, at periapsis (apoapsis), the
ADM density increases (decreases) and, thus, increases (decreases) the baryonic
sector in the subparsec region near the SMBH at the GC. As a
result, the variability of the ADM density and the baryonic matter
density are determined by the variability of the gravitational potential around
the BH, which is controlled by tidal interactions with other galaxies
in the form of the Virgo-like cluster (see e.g. \cite{Semczuk2016}).

\begin{figure}[tbp!]
  \begin{center}
    \includegraphics[width=9cm]{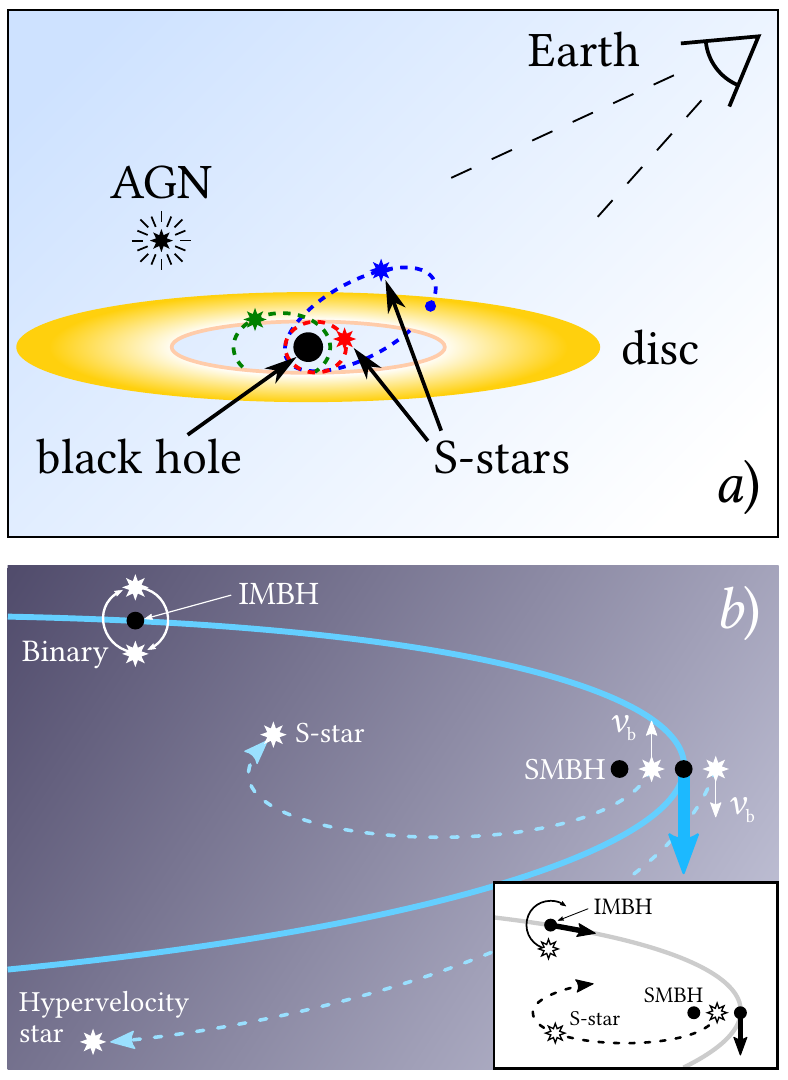}
  \end{center}
\caption{S-stars between the inner edge of the disk and the BH
``tachocline''.
\textbf{(a)} Simplified sketch of the observed variability of local positions of orbital
S-stars near the BH.
\textbf{(b)} Here is how the solution of the problem of energy conservation on the basis
of four bodies works: a single body (SMBH) exchanges partners of binary
stars orbiting around the IMBH, and through an
extreme gravitational tidal field, one star is captured (by the SMBH) and loses
energy, while the other runs away, gets all this energy and (through
hypervelocity) just flies out of the galaxy (see \cite{Hills1988,Hills1991,
BrownEtAl2005,BrownEtAl2014,BrownEtAl2018,Brown2015,Brown2016b,SubrHaas2016,
Kenyon2018,Rasskazov2019}). This is the so-called double slingshot.
\textbf{Inset:} If two objects -- an IMBH and a star rotating around it -- are
approaching a SMBH, then with three-body interactions the
gravitational tidal field can be so extreme that it can separate the star from the
IMBH. The capture or ejection of the star depends on the direction of motion of
this star relative to the pair of BHs (SMBH-IMBH). Most likely, the
star is captured by the SMBH. The resulting torus-like configuration is
determined by the Kozai-Lidova eccentric mechanism in binary SMBHs (see e.g. analogous works by
\cite{Naoz2014,Naoz2016,SubrHaas2016,Rasskazov2019}).
}
\label{fig-hypervelocity}
\end{figure}

We have found that, in contrast to the notable paper by \cite{Merritt2009},
the presence of the intermediate mass BH
(IMBH; see evidence by \cite{Takekawa2019}) orbiting inside a nuclear star
cluster at the GC can effectively help (by means of the AR-CHAIN
code \citep{Mikkola2008} and the ``indirectly observable'' ADM
halo density modulation in the center of the Milky Way) to randomize the orbits of S-stars near the SMBH,
converting the initially thin co-rotating disk (see the warped or
mini-disk in
Fig.~26c in~\cite{RusovArxiv2019}; see also \cite{ChenAmaro2014,ChenAmaro2015}) into the
almost isotropic distribution of stars moving on eccentric orbits around the
SMBH (see Fig.~\ref{fig-hypervelocity}). Here the word ``almost'', as we
believe, practically overturns the essence of understanding of the physics of
randomizing the orbits of S-stars near the SMBH, since the initial distribution of
stars is somewhat ad hoc, and the evolution to the distribution of thermal
eccentricity itself occurs at the even more special time scale.

First of all, the randomization of the orbital planes requires
$\leqslant 20~Myr$ (see \cite{Subr2019}) if the IMBH mass is
$(3.2 \pm 0.6) \times 10^4 ~M_{Sun}$ (see \cite{Takekawa2019}) and if the
orbital eccentricity is $\sim 0.7$ or greater. So, this means that in our view
of the S-stars randomization, the final distribution of the main semiaxes of star
orbits does not depend on the estimated size of the IMBH orbit
(see \cite{Merritt2009}) or, for example, on the special orientation of the
binary orbit (see \cite{Rasskazov2017}), but depends on the ``indirectly observable''
density of the DM halo at the center of the Milky Way.
Its time scale modulation is
determined by the time scale of the S-stars periods
through the
variability of the gravitational potential at the GC, which is
controlled by means of the ``observed'' disk wind rates variations or,
equivalently, the accretion rate variations of the BH.

Let us now return to the question posed above. If we take into account the
evolution of the SMBH-IMBH binaries through the case of the ejection of
high-velocity S-stars (Fig.~\ref{fig-hypervelocity}a) and B-type hypervelocity
stars (Fig.~\ref{fig-hypervelocity}b), i.e. the so-called IMBH slingshot,
then how is the time scale of the S-stars periods
(through the
variability of the gravitational potential at the center of the galaxy) driven
and controlled by the ``observable'' variations of the disk wind speeds or,
equivalently, the variations of the BH accretion rate?

It works as follows. Among the chaotically oriented orbits of S-stars, some of
them which are oriented near the fundamental plane of the GC
have a direction along the accretion to the SMBH. When an
elliptically orbital star moves from the BH periapsis (at the high
speed the capture of ADM is almost zero) to apoapsis (when the speed is
lower, the capture of ADM is not too small), it means that the
point of apoapsis is determined by the condition of hydrostatic balance, at
which the deceleration of the S-star must be identical (in the absolute value) to
the deceleration of the disk wind, or equivalently, the lower BH
accretion rate. Conversely, the increase in the speed of an elliptical-orbit
star from apoapsis to periapsis assumes the higher speed of the disk wind, or
equivalently, the higher rate of the BH accretion. As a result, the periods of
variability in the disk wind speed or accretion rate are an indicator of the
ADM variability, and consequently, an indicator of the periods of
S-stars, e.g. S102 and S2 with periods of about 11 and 16 years
(see Fig.~\ref{fig-stellar-orbits}a,b).

It also means that among the isotropically distributed S-stars there are such
stars that, although they have random orientations, do not have random
velocities and periods moving along eccentric orbits near the fundamental plane
of the GC (see Fig.~\ref{fig-hypervelocity}a).

A unique forecast of our model is the fact that the periods, velocities and
modulations of S-stars are a fundamental indicator of the modulation of the ADM halo density
in the fundamental plane
at the center of the Galaxy, which closely correlates with
the density modulation of the baryon matter near the SMBH. If the modulations of the
ADM halo at the GC lead to modulations of the ADM
density on the surface of the Sun (through vertical density waves
from the disk to the solar neighborhood), then there is an ``experimental''
anticorrelation identity of the ADM density modulation in the solar interior
and the number of sunspots.
Therefore, this is also true for the correlation identity
of the periods of S-star cycles and the sunspot cycles
(see ``experimental'' anticorrelated data in Fig.~\ref{fig-stellar-orbits}a,b).

\section{Summary and Outlook}
\label{sec-summary}

By means of photons of axion origin, we obtain three main remarkable results.

$\bullet$ First, the solar luminosity of the corona is practically identical to the 
luminosity of the photosphere (which is confirmed by 
Eqs.~(\ref{eq7.5-sep-18})-(\ref{eq7.5-sep-19})),
since the number of photons of
axion origin, which exit through the magnetic flux tubes of the photosphere,
despite the change in their density, is almost equal to the number of photons
in the low-density ambient plasma of the corona. As a result, the experimental
ratio $L_{corona}^X / L_{Sun}$ (see Fig.~\ref{fig-corona-compton}a) is the same
as the theoretical ratio $L_a^{*} / L_{Sun} \times P_{\gamma}$ (see e.g.
Eqs.~(\ref{eq7.5-sep-18})-(\ref{eq7.5-sep-19}))
in the minimum and maximum solar luminosity.

$\bullet$ Second, this means that the average energy 
($\langle E_a \rangle \approx 4.2 ~keV$ \citep{Raffelt1986}) of the photon of
axion origin can generate a temperature of the order of 
$T_a \sim 1.11 \cdot 10^7 ~K$ (see e.g.  Fig.~\ref{fig-corona-compton}a,c)
under certain conditions of coronal substances, which is close to the
temperature $T_{core} \sim 1.55 \cdot 10^7~K$ of the solar core! 
As a result, the free energy accumulated by the photons of axion origin in a
magnetic field by means of degraded spectra due to multiple Compton scattering (see Fig.~\ref{fig06}b,c),
is quickly released
and converted into heat and plasma motion with a temperature of
$\sim 4 \cdot 10^6~K$ at maximum and $\sim 1.5 \cdot 10^6~K$ at minimum of solar
luminosity.
It should be remembered that under some conditions, photons of axion origin
with very powerful flares in the corona can partially generate energy,
which corresponds to a temperature of the
order of $T_a \sim 10~MK$ (see e.g. Fig.~\ref{fig-corona-compton}b,d).

$\bullet$ Third, we are interested in the existence of a dark matter chronometer hidden
deep in the core of the Sun. The main reason is not the solar dynamos, which do
not exist due to the strong magnetic fields, but the gravitational capture of
asymmetric dark matter in the interior of the Sun. It depends on the 
holographic Babcock-Leighton mechanism as a consequence of the fundamental
holographic principle of quantum gravity in the Sun.

A unique result of our model is that the physics of ADM halo
modulation in the GC, which correlates with the modulation of the baryonic
matter density near the supermassive black hole, leads to the ADM halo density
modulations on the surface of the Sun (through the vertical density waves from
the disk to the solar neighborhood). That is the ``experimental''
anticorrelation relationship between the ADM density modulation in the interior
of the Sun and the number of sunspots. So it is indirectly proved that the ADM
halo modulation in the GC is an indicator of the periods of S-stars,
e.g. S102 with period of about 11 years, which are directly related to the
identical periods of S-star cycles and the sunspot cycles.

This means that the S-star periods (see Fig.~\ref{fig-stellar-orbits}),
starting from S102 with a period of about 11 years, S2 with period of about 16
years, S38 with period of about 22 years, S21 with period of about 30 years,
and up to S4, S9, S12, S13, S17, S18, S31 with periods of about 60 years, are
the indicators of the anti-correlation of periods of ADM density modulation
inside the Sun, which, on the other hand, control the luminosity cycles of the
Sun or sunspots.

But the main question is that we know the periods of the ADM density modulation
inside the Sun, but, unlike the assumed maximum 
($(L_{ADM})_{-} = 1.2~\cdot~10^{29}~erg/s$) and minimum 
($(L_{ADM})_{+} = 1.2~\cdot~10^{30}~erg/s$) of solar luminosity, today we do
not know how to calculate the 11-year ADM density modulation inside the Sun,
which controls the modulation of solar luminosity.

In order to understand the physics of the 11-year modulation of the solar
luminosity, we need a real interpretation of the energy transfer in the solar
interior during the scattering of dark matter. However, taking into account the
limits of significant indirect signals for detecting the results of an
asymmetric dark mother with long-range interactions, where the residual
annihilation of DM is enhanced by the Sommerfeld effect (see
e.g.~\cite{Baldes2018}), it is necessary to obtain the best solution of
spin-dependent scattering $v^2$ with a cross section of $\sigma _0~cm^2$ (at a
speed of $v_0 = 220~km/s$) and dark matter with a mass of about 5~GeV. 
This is a nontrivial result, first noted in \cite{Baldes2017}, which deserves
further study.

\appendix
\numberwithin{figure}{section}
\renewcommand{\thefigure}{\Alph{section}.\arabic{figure}}

\section{Thermomagnetic Ettingshausen-Nernst effect of toroidal magnetic field in the tachocline}
\label{appendix-a}

It is known~\citep{Schwarzschild1958} that the temperature dependence of the
rate of a thermonuclear reaction is proportional to $T^{4.5}$ in the $10^{7}~K$
region. This means that there is a clear boundary between the hot region,
which covers thermonuclear reactions and, as a consequence, high-energy photons
(gamma rays) from the core to the outer edge of the radiative zone, and a
colder region in which the dominant transport process (high-energy photons)
changes to convection. This boundary between the radiative zone and the
convection zone is the so-called tachocline. Because of the great temperature
gradient in the tachocline, the thermomagnetic 
EN~effect~\citep{Ettingshausen1886,Sondheimer1948,Spitzer1956,Kim1969} produces
strong electric currents screening the intense magnetic fields
$\sim 5 \cdot 10^7$~G of the solar core~\citep{Fowler1955,Couvidat2003}.

The thermomagnetic current can be generated in the magnetized plasma under the
quasi-steady magnetic field in the weak collision approximation (the collision
frequency much less than the positive ion cyclotron
frequency)~\citep{Spitzer1962,Spitzer2006}. For the fully ionized plasma the
EN~effect yields the current density (see Eqs.~(5-49) and
(5-52) in~\cite{Spitzer1962,Spitzer2006}):

\begin{equation}
\vec{j} _{\perp} = \frac{3 k n_e c}{2 B^2} \vec{B} \times \nabla T
\label{eq06-01}
\end{equation}

\noindent where $n_e$ is the electron number density, $B$ is the magnetic
field, $T$ is the absolute temperature, $k$ and $c$ stand for the Boltzmann constant and the speed of 
light, respectively. When $n_e = \left[ Z / (Z+1)
\right] n$, where $n = n_e + n_i$, and $n_i = n_e / Z$ is the ion number
density for a $Z$-times ionized plasma,

\begin{equation}
\vec{j} _{\perp} = \frac{3 k n c}{2 B^2} \frac{Z}{Z+1} \vec{B} \times \nabla
T .
\label{eq06-02}
\end{equation}

It exerts a force on plasma, with the force density $F$ given by

\begin{equation}
\vec{F} = \frac{1}{c} \vec{j} _{\perp} \times \vec{B} =
\frac{3 n k}{2 B^2} \frac{Z}{Z+1} \left( \vec{B} \times \nabla T \right)
\times \vec{B}\, ,
\label{eq06-03}
\end{equation}

or with $\nabla T$ perpendicular to $\vec{B}$

\begin{equation}
\vec{F} = \frac{3 n k}{2} \frac{Z}{Z+1} \nabla T\, ,
\label{eq06-04}
\end{equation}

which leads to the magnetic equilibrium (see Eq.~(4-1)
in~\cite{Spitzer1962}):

\begin{equation}
\vec{F} = \frac{1}{c} \vec{j} _{\perp} \times \vec{B} = \nabla p
\label{eq06-05}
\end{equation}

with $p = nkT$. By equating~(\ref{eq06-04})
and~(\ref{eq06-05}),

\begin{equation}
\frac{3 n k}{2} \frac{Z}{Z+1} \nabla T = nk \nabla T + kT \nabla n
\label{eq06-06}
\end{equation}

\noindent or

\begin{equation}
a \frac{\nabla T}{T} + \frac{\nabla n}{n} = 0,
~~~ where ~~ a = \frac{2 - Z}{2(Z+1)} ,
\label{eq06-06a}
\end{equation}

\noindent we obtain the condition

\begin{equation}
T ^a n = const .
\label{eq06-07}
\end{equation}

For the singly ionized plasma with $Z=1$,

\begin{equation}
T ^{1/4} n = const .
\label{eq06-08}
\end{equation}

For the doubly ionized plasma ($Z=2$) $n=const$. Finally, in the limit of large 
$Z$, $T^{-1/2}n = const$, and $n$ does not depend on $T$
strongly, as opposed to the case of the plasma at a constant pressure with
$Tn=const$. Thus, the thermomagnetic currents can change the pressure
distribution in the magnetized plasma considerably.

Choosing the Cartesian coordinate system with $z$ axis along $\nabla T$,
$x$ axis along the magnetic field and $y$ axis along the current,
and assuming the fully ionized hydrogen plasma with $Z=1$ in the tachocline,
we obtain

\begin{equation}
{j} _{\perp} = {j} _y = - \frac{3 n k c}{4 B} \frac{dT}{dz}. 
\label{eq06-09}
\end{equation}

From Maxwell's equation $4 \pi \vec{j}_{\perp}/ c = \nabla \times \vec{B}$, one has

\begin{equation}
{j} _y = \frac{c}{4 \pi} \frac{dB}{dz}, 
\label{eq06-10}
\end{equation}

\noindent
then by equating~(\ref{eq06-09}) and~(\ref{eq06-10}) we derive

\begin{equation}
2B \frac{dB}{dz} = -6 \pi k n \frac{dT}{dz}.
\label{eq06-11}
\end{equation}

From~(\ref{eq06-08}) one has

\begin{equation}
n = \frac{n _{tacho} T_{tacho}^{1/4}}{T^{1/4}},
\label{eq06-12}
\end{equation}

\noindent where the values $n = n_{tacho}$ and $T = T_{tacho}$ correspond to
the overshoot tachocline. Substituting~(\ref{eq06-12}) into~(\ref{eq06-11}), we
find

\begin{equation}
d\left(B^2\right) = -\frac{6 \pi k n _{tacho} T_{tacho}^{1/4}}{T^{1/4}} dT .
\label{eq06-13}
\end{equation}

\begin{figure}[tb]
\begin{center}
\includegraphics[width=8cm]{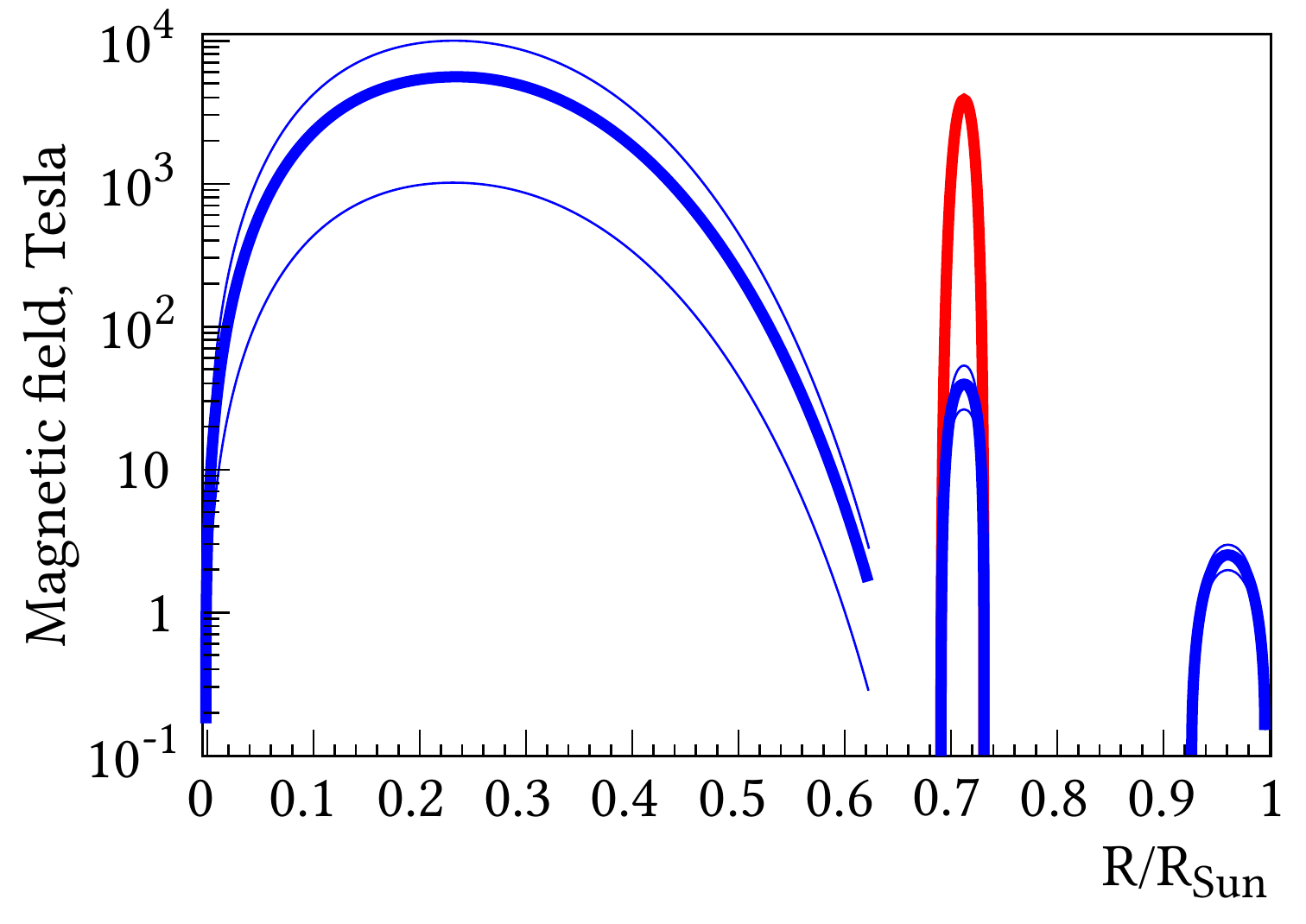}
\end{center}
\caption{The reconstructed solar magnetic field (in blue) simulation
from~\cite{Couvidat2003}: 10$^3$-10$^4$~Tesla (left), 30-50~Tesla (middle) and
2-3~Tesla (right), with temperature of $\sim$9~MK, $\sim$2~MK
and~$\sim$200~kK, respectively. The thin lines show the estimated range of
values for each magnetic field component. Internal rotation was not included in
the calculation. An additional axion production can modify both
intensity and shape of the solar axion spectrum (Courtesy Sylvaine 
Turck-Chi\`{e}ze (see Fig.~2 in~\cite{Zioutas2007})). The reconstructed solar 
magnetic field (in red) simulation from~(\ref{eq06-16}): $4 \cdot 10^3$~T in 
tachocline ($\sim0.7 R_{Sun}$).} 
\label{fig-R-MagField}
\end{figure}

\noindent As a result of integrating in the limits $[B_{tacho},0]$ on the
left and $[0,T_{tacho}]$ on the right,

\begin{equation}
\frac{B_{tacho}^2}{8 \pi} = n _{tacho} kT_{tacho}\, ,
\label{eq06-14}
\end{equation}

\noindent
expressing the fact that the magnetic field of the thermomagnetic current in
the overshoot tachocline ``neutralizes'' the magnetic field of the solar 
interior completely, so that the projections of the magnetic field in the
tachocline and in the core are equal but of opposite directions 
(see Fig.~\ref{fig-R-MagField}).

An intriguing question arises here of what forces are the cause of the
shielding of the strong magnetic fields of the solar core and the radiation
zone, which would be related to the enormous magnetic pressure in the overshoot
tachocline,

\begin{equation}
\frac{B_{tacho}^2}{8 \pi} = p_{ext} \approx 6.5 \cdot 10^{13} \frac{erg}{cm^3} ~~
at ~~ 0.7 R_{Sun}, 
\label{eq06-15}
\end{equation}

\noindent
where the gas pressure $p_{ext}$ in the solar tachocline ($\rho
\approx 0.2 ~g\cdot cm^{-3}$ and $T \approx 2.3 \cdot 10^6 K$
\citep{Bahcall1992} at~$0.7 R_{Sun}$) yields the toroidal magnetic field

\begin{equation}
B_{tacho} \simeq 4100 ~T = 4.1 \cdot 10^7 ~G\, .
\label{eq06-16}
\end{equation}

Propagating the question above even further, one may ask: ``What kind of
mysterious nature of the solar tachocline gives birth to the repulsive
magnetic field through the thermomagnetic EN~effect?''
And finally, ``If the thermomagnetic EN~effect exists in the
tachocline, what is its physical nature?''
The essence of this physics will be discussed below.

\section{Almost empty magnetic flux tube nearby of the tachocline and physics of the mean free path of the axion-origin photon}
\label{appendix-b}

It is known that the unsolved problem of energy transport by magnetic flux
tubes at the same time represents another unsolved problem related to the
sunspot darkness (see 2.2 in \cite{Rempel2011}). Of all the known concepts 
playing a noticeable role in understanding the connection between the energy
transport and sunspot darkness, let us consider the most significant theory,
in our view. It is based on the Parker-Biermann cooling
effect~\cite{Parker1955a,Biermann1941,Parker1979b} and originates from the
early works of Biermann \cite{Biermann1941} and Alfv\'{e}n \cite{Alfven1942}.

As you know, the Parker-Biermann cooling
effect~\cite{Parker1955a,Biermann1941,Parker1979b}, which plays a role in our
current understanding, originates from Biermann~\cite{Biermann1941} and
Alfv\'{e}n~\cite{Alfven1942}: in a highly ionized plasma, the electrical
conductivity can be so large that the magnetic fields are frozen into the
plasma. Biermann realized that the magnetic field in the spots themselves can
be the cause of their coolness -- it is colder because the magnetic field
suppresses the convective heat transfer. Hence, the darkness of the spot is
due to a decrease in surface brightness.

Parker~\cite{Parker1955a,Parker1974a,Parker1974b,Parker1974c,Parker1979b} 
has pointed out that the magnetic field can be compressed to the enormous
intensity only by reducing the gas pressure within the flux tube relative to
the pressure outside, so that the external pressure compresses the field.
The only known mechanism for reducing the internal pressure sufficiently is a
reduction of the internal temperature over several scale heights so that the
gravitational field of the Sun pulls the gas down out of the tube (as described
by the known barometric law $dp / dz = - \rho g$). Hence it appears that the
intense magnetic field of the sunspot is a direct consequence of the observed
reduced temperature~\cite{Parker1955a}.

On the other hand, Parker~\cite{Parker1974c,Parker1977} has also pointed out
that the magnetic inhibition of convective heat transport beneath the sunspot,
with the associated heat accumulation below, raises the temperature in the
lower part of the field. The barometric equilibrium leads to enhanced gas
pressure upward along the magnetic field, causing the field to disperse rather
than intensify. Consequently, Parker~\cite{Parker1974c} argued that the
temperature of the gas must be influenced by something more than the inhibition
of heat transport!

Currently it is believed (see~\cite{Perri2018}) that the ultimate
goal is to simulate the formation of sunspots, where the convective heat
transfer is either suppressed by a magnetic field~\cite{Biermann1941} or the
cooling is enhanced~\cite{Parker1974a}. Thus, in general, it seems that we
do not have a satisfactory explanation for the sunspot phenomenon yet. The 
question of the lack of energy flow or cooling, taking into account the
magnetic field strength, remains open. According to \cite{Perri2018},
this will be another urgent goal for further research.

\begin{figure*}[tbp!]
\begin{center}
\includegraphics[width=15cm]{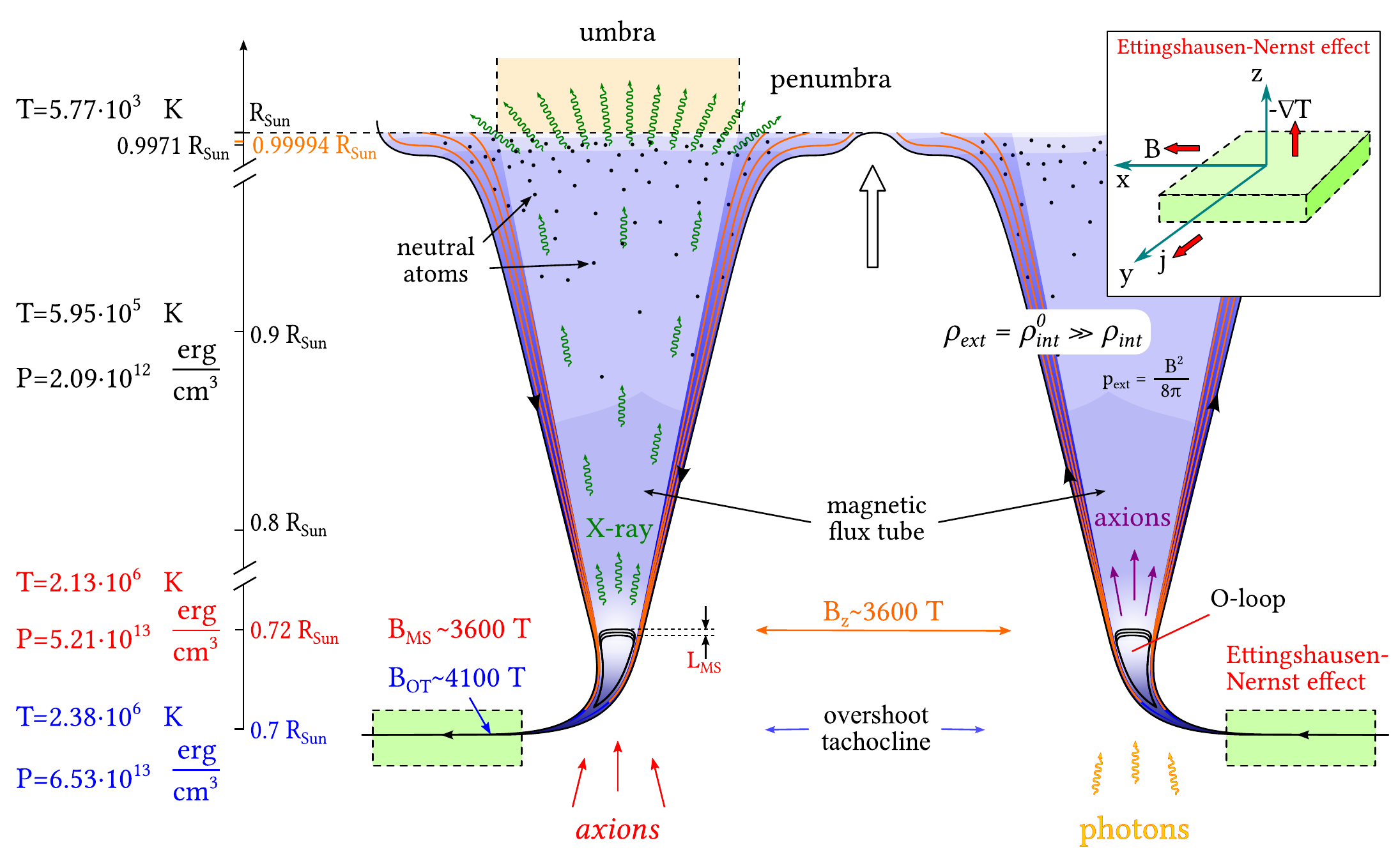}
\end{center}
\caption{Topological effects of magnetic reconnection inside the
magnetic tubes with the ``magnetic steps'' (see Fig.~6 in Rusov~et~al.~\cite{RusovArxiv2019}). 
The left panel shows the
temperature and pressure change along the radius of the Sun from the tachocline
to the photosphere \citep{Bahcall1992}, $L_{MS}$ is the height of the magnetic
shear steps. At $R \sim 0.72~R_{Sun}$ the vertical magnetic field reaches $B_z
\sim 3600$~T, and the magnetic pressure $p_{ext} = B^2 / 8\pi 
\simeq 5.21 \cdot 10^{13}~erg/cm^3$ \citep{Bahcall1992}.
Then a horizontal magnetic field is developed through the well-known Kolmogorov
turbulent cascade (see Fig.~\ref{fig-Kolmogorov-cascade}).
The very cool regions along the entire convective zone caused by the
Parker-Biermann cooling effect, have the virtually zero internal gas pressure
because of the maximum magnetic pressure inside the magnetic tubes.}
\label{fig-lampochka}
\end{figure*}

Our unique alternative idea is that the explanation of sunspots is
based not only on the suppression of convective heat transfer by a strong
magnetic field~\cite{Biermann1941}, but, in contrast to the enhanced 
cooling~\cite{Parker1974a}, the Parker-Biermann cooling effect appears as a result
of the disappearance of barometric equilibrium~\cite{Parker1974a}, which is
confirmed by axions of photonic origin from the photon-axion oscillations in
the O-loop near the tachocline (see Fig.~\ref{fig-lampochka}).

According to~\ref{appendix-a},
the $B \sim 4100$~T magnetic field in the overshoot tachocline and the
Parker-Biermann cooling effect in an almost empty magnetic tube can produce the O-loops with the horizontal
magnetic field $B_{MS} \approx B(0.72 R_{Sun}) \sim 3600$~T stretching for about
$L_{MS} \sim 1.28 \cdot 10^4 ~km$, (see e.g. Fig.~6ab in \cite{RusovArxiv2019}),
and surrounded by virtually zero internal gas pressure of the magnetic tube
(see Fig.~6a in \cite{RusovArxiv2019}).
As an example, let us show Fig.~\ref{fig-lampochka} and 
$p_{ext} = B_{0.72R_{Sun}}^2 / 8\pi = 5.21 \cdot 10^{13} ~erg/cm^3$~\cite{Bahcall1992}.

\begin{figure*}[tbp!]
\begin{center}
\includegraphics[width=16cm]{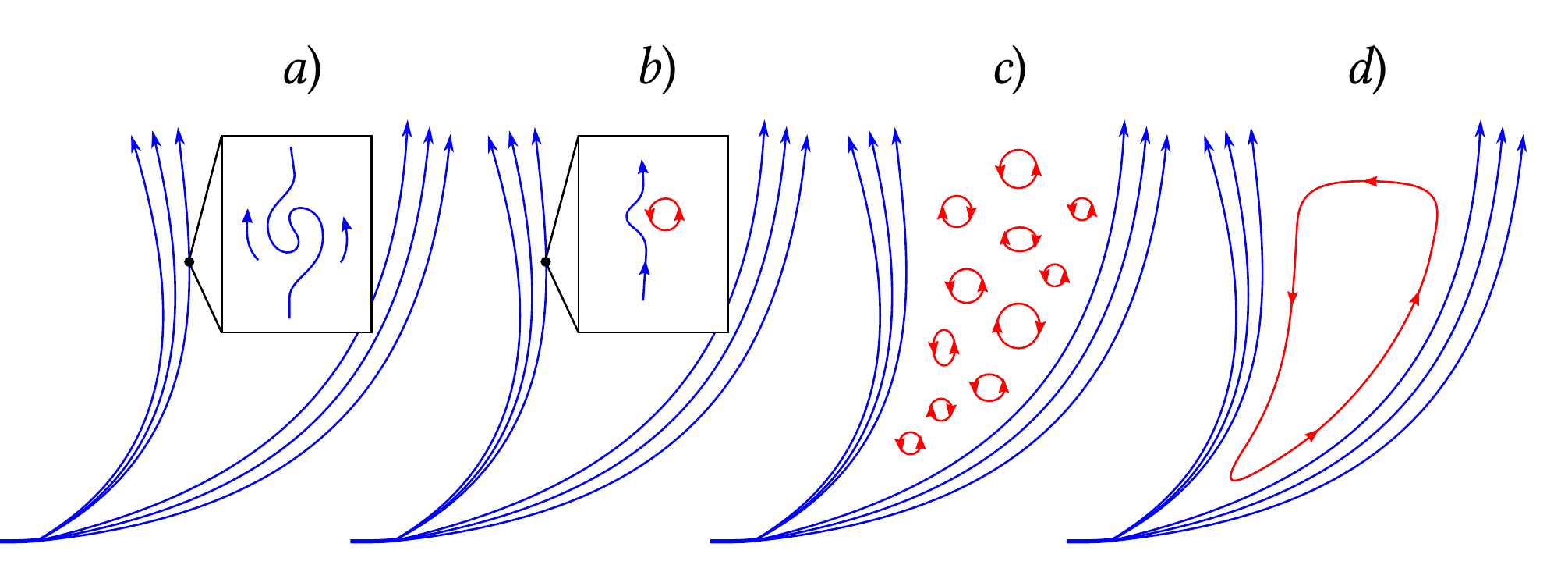}
\end{center}
\caption{Kolmogorov turbulent cascade
\citep{Kolmogorov1941,Kolmogorov1968,Kolmogorov1991} and magnetic reconnection
in the lower layers inside a unipolar magnetic tube. Common to these various
turbulent systems is the presence of the inertial range of Kolmogorov, through
which the energy is cascaded from large to small scales, where dissipative
mechanisms (as a consequence of magnetic reconnection) overcome the turbulent
energy in plasma heating.}
\label{fig-Kolmogorov-cascade}
\end{figure*}

On the other hand, we showed that the topological effects of magnetic
reconnection inside magnetic tubes near the tachocline form the O-loop
(Fig.~\ref{fig-Kolmogorov-cascade}) through the Kolmogorov turbulent cascade
(see Fig.13 in~\cite{RusovArxiv2019}). Their ``magnetic steps'' $L_{MS}$
participate in the formation of photons of axion origin (see 
Fig.~\ref{fig-lampochka}).

This rises another beautiful problem which is associated with our problem of
almost total suppression of radiative heating in virtually empty magnetic
tubes (see Fig.~\ref{fig-lampochka}). Let us remind that the high-energy photons going from the
radiation zone through the horizontal field of the O-loop near the tachocline
(Fig.~\ref{fig-lampochka}) are turned into
axions, thus almost completely eliminating the radiative heating in the
virtually empty magnetic tube. Some small photon flux can still pass
through the ``ring'' between the O-loop and the tube walls and reach the
penumbra (Fig.~\ref{fig-lampochka}).

Thus, the remarkable problem is that, on the one hand, there is a small flux of
photons coming from the overshoot tachocline, which passes through the ``ring''
of a strong magnetic tube, which by means of convective heating physics
$(dQ/dt)_2$ (see Fig.~\ref{fig-GalacticDisk-MagTube}; 
Sect. 3.1.3.1 in~\cite{RusovArxiv2019}) allows one
to determine both the lifetime of the magnetic flux tube from the tachocline to
the surface of the Sun, and the reconnection rate  $V_{rec}$, which
is determined only by the process of magnetic loop raising and consequent
sunspot vanishing from the surface of the Sun!

Therefore, as we understand, the existence of a magnetic O-loop (see 
Fig.~\ref{fig-Kolmogorov-cascade}) and the Parker-Biermann cooling effect
(Fig.~\ref{fig-lampochka}), which completely suppresses convection in an
almost empty magnetic tube, is the cause of the sunspot darkness.

Finally, let us show the theoretical estimates of the Rosseland mean opacity
and the axion-photon oscillations in the magnetic flux tubes.

It is essential to find the physical solution to the problem of solar convective zone which would fit the opacity experiments. The full calculation of solar opacities, which depend on the chemical composition, pressure and temperature of the gas, as well as the wavelength of the incident light, is a complex endeavor. The problem can be simplified by using the mean opacity averaged over all wavelengths, so that only the dependence on the gas physical properties remains (see e.g. \cite{Rogers1994,Ferguson2005,Bailey2009}). The most commonly used is the Rosseland mean opacity $k_R$, defined as:

\begin{equation}
\frac{1}{k_R} = \left. \int \limits_{0}^{\infty} d \nu \frac{1}{k_\nu} \frac{dB_\nu}{dT} \middle/ 
\int \limits_{0}^{\infty} d \nu \frac{dB_\nu}{dT} \right.\, ,
\label{eq06v2-02}
\end{equation}

\noindent
where $dB_\nu / dT$ is the derivative of the Planck function with respect to 
temperature, $k_{\nu}$ is the monochromatic opacity at frequency $\nu$ of the 
incident light or the total extinction coefficient,  including stimulated 
emission plus scattering. A large value of the opacity indicates strong 
absorption from beam of photons, whereas a small value indicates that the beam 
loses very little energy as it passes through the medium.

\begin{figure}[tbp!]
\begin{center}
\includegraphics[width=15cm]{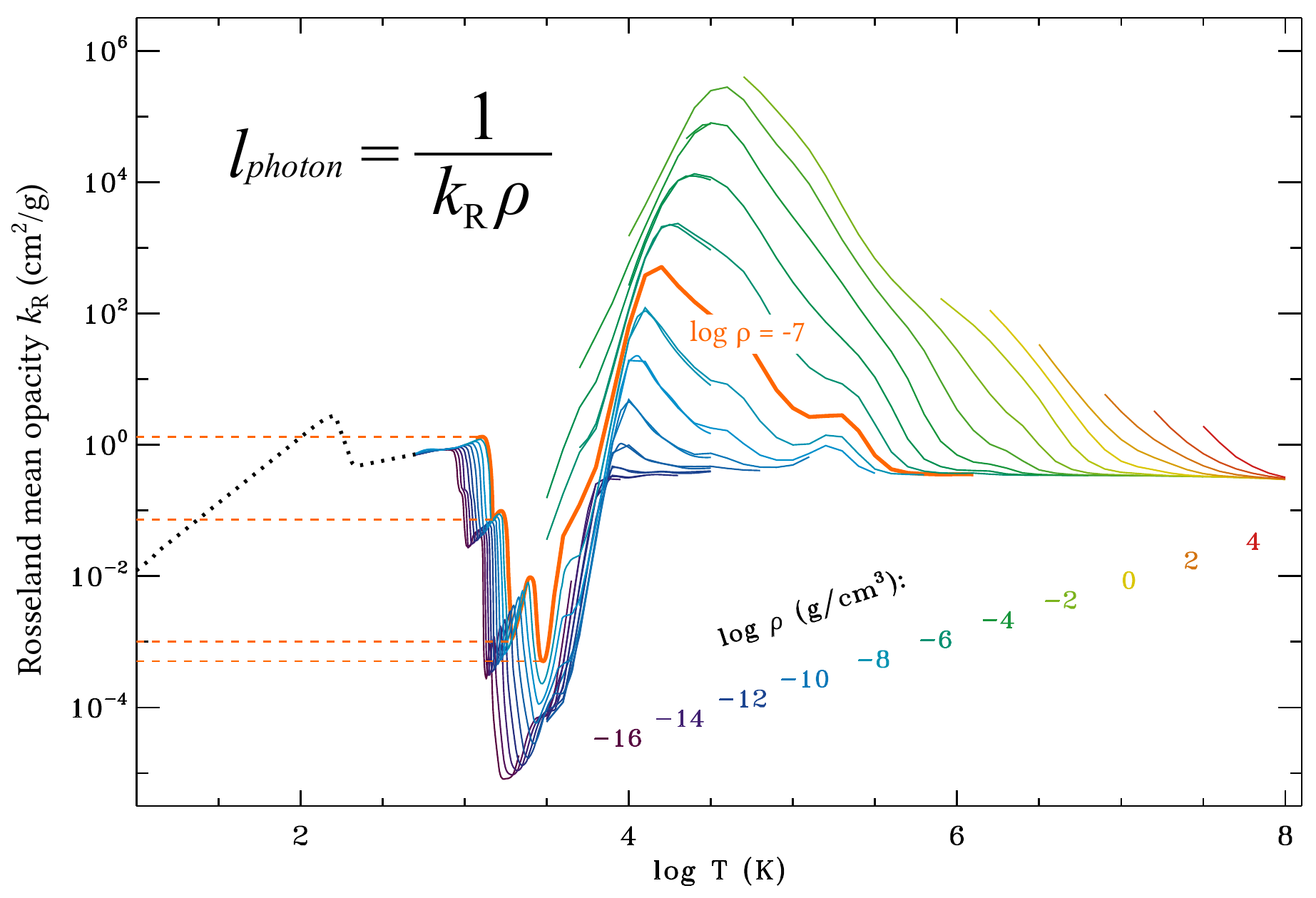}
\end{center}
\caption{Rosseland mean opacity $k_R$, in units of $cm^2 g^{-1}$, shown versus 
temperature (X-axis) and density (multi-color curves, plotted once per decade),
computed with the solar metallicity of hydrogen and helium mixture X=0.7 and 
Z=0.02. The panel shows curves of $k_R$ versus temperature for several 
``steady'' values of density, labeled by the value of $\log {\rho}$ (in 
$g/cm^3$). Curves that extend from $\log {T} = 3.5$ to 8 are from the Opacity 
Project (opacities.osc.edu). Overlapping curves from $\log {T} = 2.7$ to 4.5 
are from \cite{Ferguson2005}. The lowest-temperature region (black dotted 
curve) shows an estimate of ice-grain and metal-grain opacity from 
\cite{Stamatellos2007}. Adopted from \cite{Cranmer2015}.}
\label{fig-opacity}
\end{figure}

Note that the Rosseland opacity is the harmonic mean, in which the greatest 
contribution comes from the lowest values of opacity, weighted by a function 
that depends on the rate at which the blackbody spectrum varies with 
temperature (see Eq.~(\ref{eq06v2-02}) and Fig.~\ref{fig-opacity}), and the
photons are most efficiently transported through the ``windows'' where $k_\nu$ 
is the lowest (see Fig.~2 in \cite{Bailey2009}).

Taking the Rosseland mean opacities shown in Fig.~\ref{fig-opacity}, one may 
calculate, for example, four consecutive cool ranges within the convective 
zone (Fig.~\ref{fig-lampochka}), where the internal gas pressure $p_{int}$ is 
defined by the following values:

\begin{equation}
p_{int} = n k T, ~where~ 
\begin{cases}
T \simeq 10^{3.48} ~K, \\
T \simeq 10^{3.29} ~K, \\
T \simeq 10^{3.20} ~K, \\
T \simeq 10^{3.11} ~K, \\
\end{cases}
\rho = 10^{-7} ~g/cm^3\, .
\label{eq06v2-03}
\end{equation}

Since the inner gas pressure~(\ref{eq06v2-03}) decreases towards the surface of
the Sun, so that

\begin{align}
p_{int} &(T = 10^{3.48} ~K) \vert _{\leqslant 0.85 R_{Sun}}  > 
p_{int} (T = 10^{3.29} ~K) \vert _{\leqslant 0.9971 R_{Sun}} > \nonumber \\
& > p_{int} (T = 10^{3.20} ~K) \vert _{\leqslant 0.99994 R_{Sun}} > 
p_{int} (T = 10^{3.11} ~K) \vert _{\leqslant R_{Sun}} ,
\label{eq06v2-04}
\end{align}

\noindent
it becomes evident that the neutral atoms appearing in the upper convection 
zone ($\geqslant 0.85 R_{Sun}$) cannot descend deep to the base of the 
convection zone, i.e. the tachocline (see Fig.~\ref{fig-lampochka}).

Therefore, it is very important to examine the connection between the Rosseland 
mean opacity and axion-photon oscillations in a magnetic flux tube.

\begin{figure}[tb]
  \begin{center}
    \includegraphics[width=15cm]{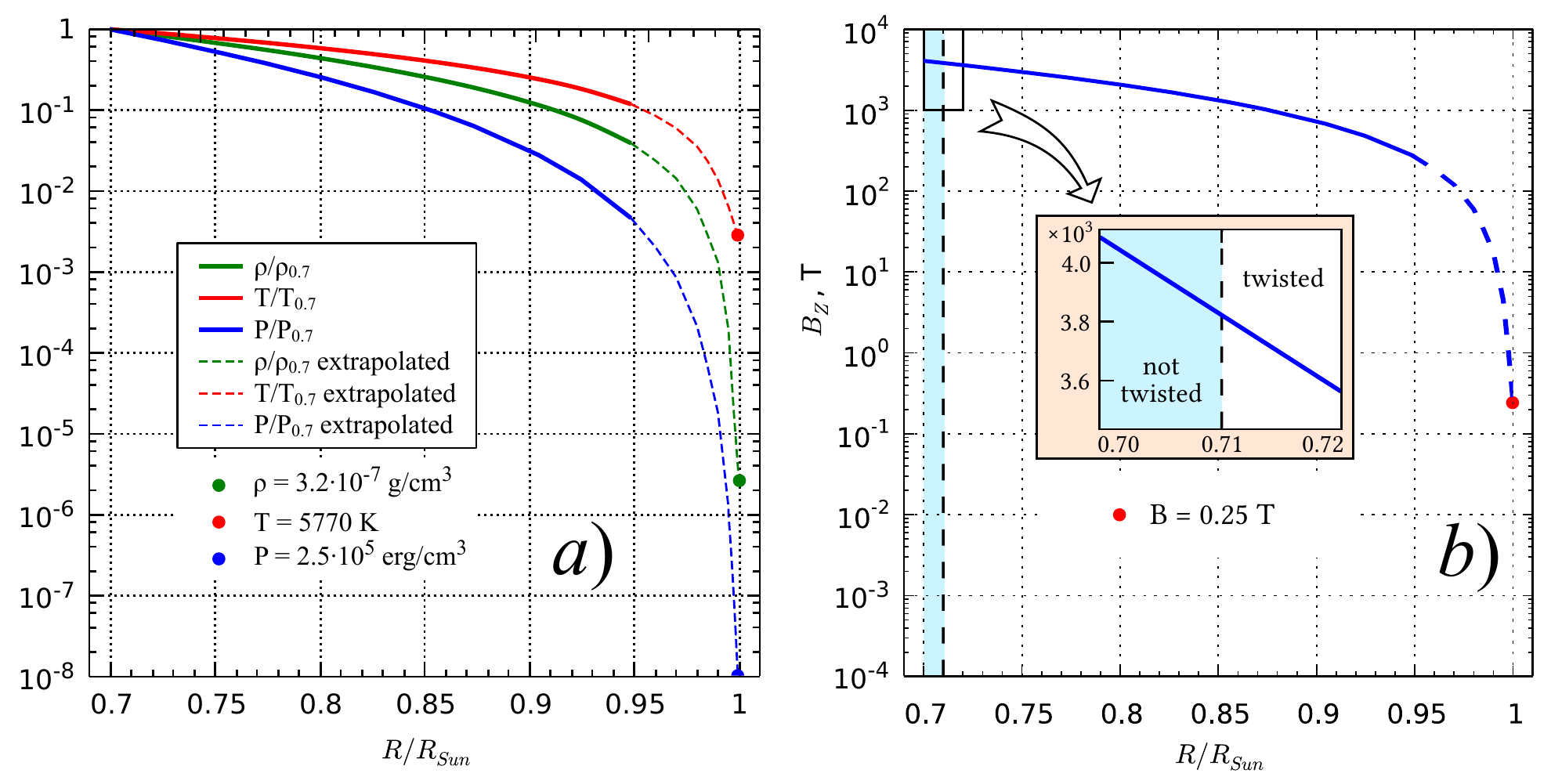}
  \end{center}
\caption{
(a) Normalized external temperature, density and gas pressure as functions of 
the solar depth $R/R_{Sun}$. The standard solar model with $He$ diffusion 
\citep{Bahcall1992} was used  for $R < 0.95 R_{Sun}$ (solid lines). The dotted 
lines mark extrapolated values.
(b) Variation of the magnetic field strength $B_z$  along the emerging
$\Omega$-loop as a function of the solar depth $R / R_{Sun}$ throughout the
convection zone. The solid blue line marks the permitted values for
the same standard solar model with $He$ diffusion \citep{Bahcall1992} starting at
the theoretical estimate of the magnetic field 
$B_{tacho} \approx 4100$~T. The dashed line is the continuation, 
according to the existence of the very cool regions inside the magnetic tube. 
The red point marks the up-to-date observations showing the mean magnetic field 
strength at the level $\sim 0.25~T = 2500 ~G$ \citep{Pevtsov2011,Pevtsov2014}.}
\label{fig-Bz}
\end{figure}

In this regard let us consider the superintense magnetic $\Omega$-loop
formation in the overshoot tachocline through the local shear caused by the
high local concentration of azimuthal magnetic flux. The buoyant force
acting on the $\Omega$-loop decreases slowly with concentration, so the vertical
magnetic field of the $\Omega$-loop reaches $B_z \sim 3600$~T at about 
$R / R_{Sun} \sim 0.72$ (see Fig.~\ref{fig-lampochka} and Fig.~\ref{fig-Bz}b).
Because of the magnetic pressure (see Fig.~\ref{fig-lampochka}) 
$p_{ext} = B_{0.72 R_{Sun}}^2 / 8\pi = 5.21\cdot 10^{13}~erg/cm^3$
\citep{Bahcall1992}, this leads to significant cooling of the $\Omega$-loop
tube (see Fig.~\ref{fig-lampochka}).

Eventually we suppose that the X-rays, through the axion-photon oscillations
in the magnetic O-loop near the tachocline, 
channel along the ``cool'' region of the Parker-Biermann magnetic tube
(Fig.~\ref{fig-lampochka}) and effectively supply the necessary photons of axion origin
``channeling'' in the magnetic tube to the photosphere while the convective heat transfer is heavily
suppressed.

In this context it is necessary to have a clear view of the energy transport by X-rays of axion origin, which are a primary transfer mechanism. The recent improvements in the calculation of the radiative properties of solar matter have helped to resolve several long-standing discrepancies between the observations and the predictions of theoretical models (see e.g. \cite{Rogers1994,Ferguson2005,Bailey2009}), and now it is possible to calculate the photon mean free path (Rosseland length) for Fig.~\ref{fig-opacity}:

\begin{equation}
l_{photon} = \frac{1}{k_R \rho} \sim 
\begin{cases}
2 \cdot 10^{10} ~cm  ~~ & for ~~ k_R \simeq 5 \cdot 10^{-4} ~cm^2/g, \\
10^{10} ~cm          ~~ & for ~~ k_R \simeq 10^{-3} ~cm^2/g, \\
1.5 \cdot 10^{8} ~cm ~~ & for ~~ k_R \simeq 6.7 \cdot 10^{-2} ~cm^2/g, \\
10^{7} ~cm           ~~ & for ~~ k_R \simeq 1 ~cm^2/g,
\end{cases}
~~ \rho = 10^{-7} ~g/cm^3\, ,
\label{eq06v2-05}
\end{equation}

\noindent
where the Rosseland mean opacity values $k_R$ and density $\rho$ are chosen so 
that the very low internal gas pressure $p_{int}$ (see Eq.~(\ref{eq06v2-03})) 
along the entire magnetic tube almost does not affect the external gas pressure
$p_{ext}$ (see (\ref{eq06v2-04}) and Fig.~\ref{fig-opacity}).

Let us now examine the appearance of the X-rays of axion origin, induced by the
magnetic field variations near the tachocline (Fig.~\ref{fig-lampochka}), and
their impact on the Rosseland length (see~(\ref{eq06v2-04}))
inside the cool region of the magnetic tubes.

Let us remind that the magnetic field strength $B_{tacho}\sim 4100$~T in the overshoot 
tachocline (see Fig.~\ref{fig-lampochka}) and the 
Parker-Biermann cooling effect (see Fig.~\ref{fig-lampochka}) lead to the corresponding 
value of the magnetic field strength $B(0.72 R_{Sun}) \sim 3600$~T
(see Fig.~\ref{fig-lampochka}), which in its turn implies the virtually zero 
internal gas pressure of the magnetic tube.

As it is shown above, on the basis of Kolmogorov turbulent cascade 
(see Fig.~\ref{fig-Kolmogorov-cascade}), the topological effect of the
magnetic reconnection inside the $\Omega$-loop results in the formation of the
so-called O-loops (Fig.~\ref{fig-lampochka}).
It is possible to derive the value
of the horizontal magnetic field of the magnetic steps at the top of the O-loop:
$B_{MS} \approx B(0.72 R_{Sun}) \sim 3600$~T.

So in the case of the large enough Rosseland length (see Eq.~(\ref{eq06v2-05})), 
X-rays of axion origin, induced by the horizontal magnetic field in O-loops, reach
the photosphere freely, while in the photosphere itself, according to the 
Rosseland length

\begin{equation}
l_{photon} \approx 100 ~km < l \approx 300 \div 400 ~km,
\label{eq06v2-06}
\end{equation}

\noindent
these photons undergo multiple Compton scattering
producing a typical directional pattern
(Fig.~\ref{fig-lampochka}).

Aside from the X-rays of axion origin with mean energy of $4.2~keV$
\citep{Andriamonje2007,Zioutas2009}, there are
also $h \nu \sim 0.95 ~keV$ X-rays (originating from the interface between the
radiation zone and overshoot tachocline, according to a theoretical estimate by
\cite{Bailey2009}). Such
X-rays would produce the Compton-scattered photons with mean energy of
$\sim 0.95~keV$. These photons ``disappear'' by inverse X-ray transformation
into axions (see Fig.~\ref{fig-lampochka}). This way the $\sim 0.95~keV$ X-rays do not
contradict the known measurements of the photons with mean energy of $\sim 4~keV$ 
(see Fig.~1 in \cite{Andriamonje2007}) by involving the X-rays
of axion origin in O-loops (see Figs.~\ref{fig-lampochka}).

And finally, let us emphasize that we have just shown a theoretical possibility
of the time variation  of the  sunspot activity to correlate with the flux 
of the X-rays of axion origin; the latter being controlled by the magnetic 
field variations near the overshoot tachocline. As a result, it may be 
concluded that the axion mechanism for solar luminosity variations 
based on the lossless X-ray ``channeling'' along the
magnetic tubes allows to explain the effect of the almost complete suppression of the
convective heat transfer, and thus to understand the known puzzling darkness of
the sunspots \citep{Rempel2011}.

In addition to the energy source -- photons of axion origin in the corona,
which are the source of dark matter of axions in the core of the Sun, we are
interested in the problems of both energy propagation and energy scattering.

\begin{figure}[tb!]
\begin{center}
\includegraphics[width=16cm]{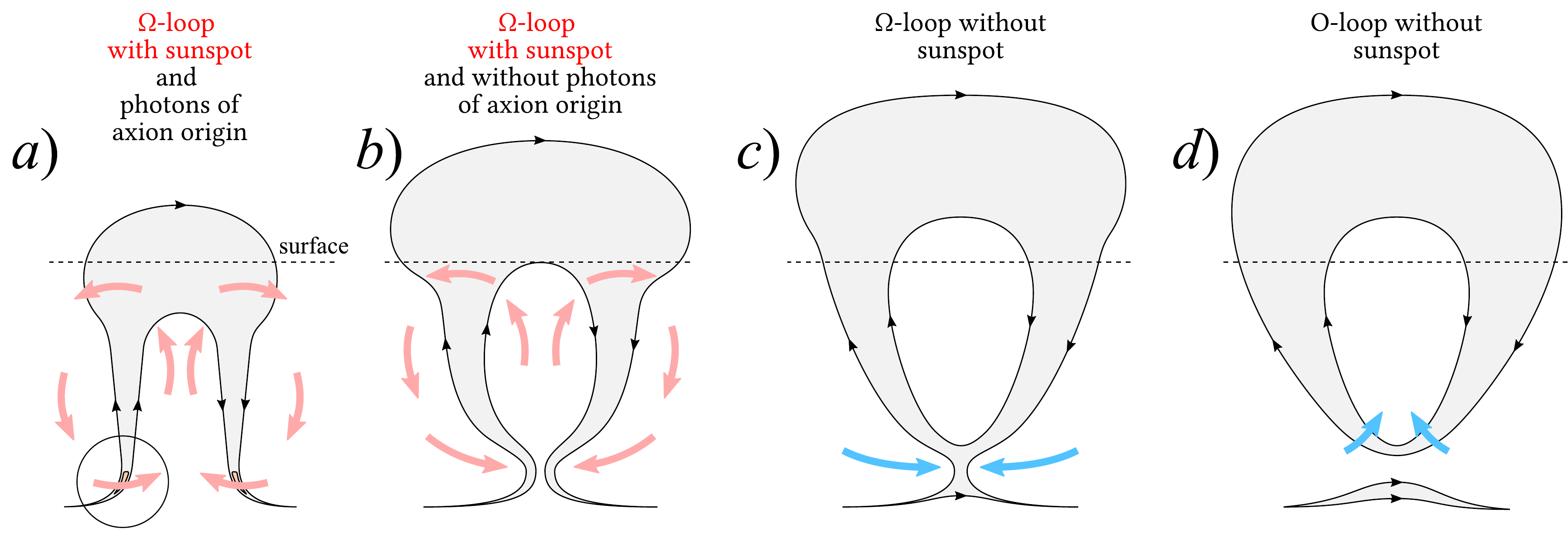}
\end{center}
\caption{A sketch of the magnetic reconnection near the tachocline.
\textbf{(a)} $\Omega$-loop forms the sunspot umbra (with photons of axion
origin) via the indirect thermomagnetic EN~effect
(\ref{appendix-a});
\textbf{(b)} $\Omega$-loop with a sunspot (without photons of axion origin);
pink arrows show the upward convective flow between the ``legs'' of the
$\Omega$-loop during its rise from the tachocline to the visible surface;
\textbf{(c)} $\Omega$-loop with reconnection and without a sunspot;
\textbf{(d)} O-loop without a sunspot.
Going through the stages (a), (b), (c), (d) (left to right), the convection
around the rising $\Omega$-loop ``closes'' it at its base, then a free O-loop is
formed via reconnection, and the initial configuration of the azimuthal field
at the bottom of this region is restored. Blue arrows show the substance motion
leading to the loop ``legs'' connection.}
\label{fig-lower-reconnection}
\end{figure}

This solution explicitly depends on the lifetime of the magnetic tubes
rising from the tachocline to the solar surface. Therefore, because of the
magnetic reconnection in the lower layers (see Fig.~4 in \cite{Parker1994}),
it is not the final stage of the simulation. The $\Omega$-loop, forming the
sunspot umbra (Fig.~\ref{fig-lower-reconnection}a) through the
convection suppression from the tachocline to the photosphere, also gives rise
to the convective upflow around the $\Omega$-loop forming the sunspot penumbra.
Because of the pre-reconnection (Fig.~\ref{fig-lower-reconnection}a,b) when the
``legs'' of the $\Omega$-loop collide (Fig.~\ref{fig-lower-reconnection}b), the
convective flow is generated at the base of the convection zone.

As we noted in Sec.~\ref{sec-corona-heating-solution}, 
the intermittent appearance of a new magnetic flux from the convective zone
(which originates from twist in flux tubes 
(see~\cite{Archontis2012,Schmieder2014,Pontieu2014}) in the corona is
the most important process for the dynamic evolution of the coronal magnetic field~\cite{Galsgaard2007,Fang2010,ArchontisHood2012,ArchontisHood2013},
in which the rearrangement of the intersection of closed coronal lines of force
causes the accumulation of coronal field strength. When the strength exceeds a
certain threshold, the stability of the magnetic field configuration is broken
and erupts in the form of nanoflares, the magnetic reconnection of which is
associated with the escaping magnetic tube from the surface of the Sun, while
the initial magnetic reconnection first occurs near the tachocline in the
convective zone (see e.g. Fig.~\ref{fig-lower-reconnection}c,d).

From here we understand that the flux of photons of axion origin, emitted from the
photosphere of an almost empty magnetic tube (see 
Fig.~\ref{fig-lower-reconnection}a), and the coronal magnetic field strength
(which arises due to the twisting of the almost empty magnetic tubes, but
without the sunspots (see
Fig.~\ref{fig-lower-reconnection}c,d)), have global and local space-time
scales in the corona respectively.

This means that the photon flux of axion origin in the corona, which, on the
basis of the plasma reaction to heating, is realized due to interaction with
electrons in the outer layers of the Sun, has a global space-time scale and is
a direct source of energy propagation, as well as the energy scattering!

\section{The fundamental holographic principle of tachocline}
\label{appendix-c}

\begin{figure*}[tbp!]
\begin{center}
\includegraphics[width=15cm]{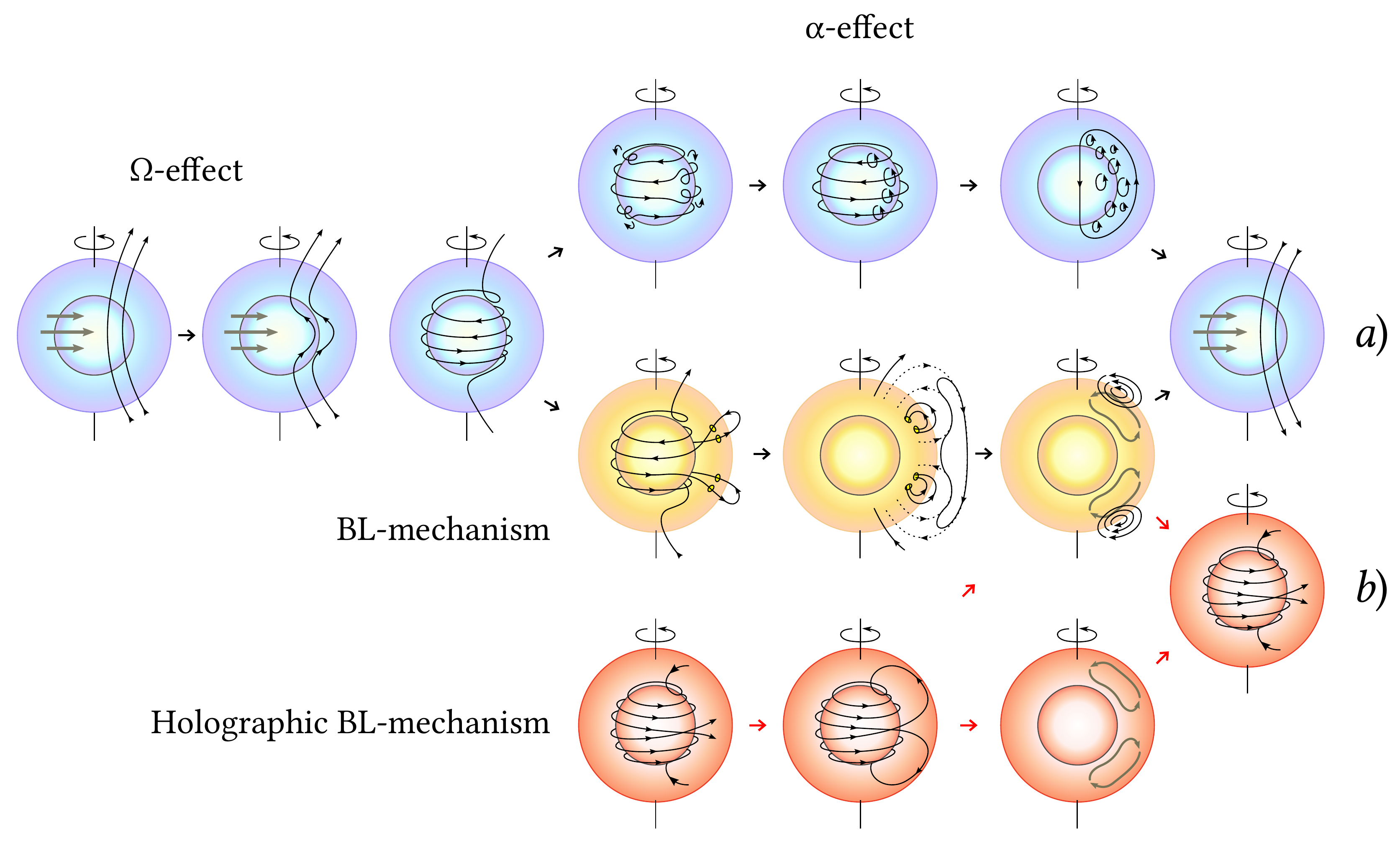}
\end{center}
\caption{An illustration of the main possible processes of a magnetically
active star of the Sun type.
(a) $\alpha$-effect, $\Omega$-effect and BL~mechanism as components of the
solar dynamo model. The $\Omega$-effect (blue) depicts the transformation of the
primary poloidal field into a toroidal field by differential rotation.
Regeneration of the poloidal field is then performed either by the
$\alpha$-effect (top) or by the BL~mechanism (yellow in the
middle). In case of $\alpha$-effect, the toroidal field at the base of the
convection zone is subject to cyclonic turbulence. In the BL~mechanism, the
main process of regeneration of the poloidal field (based on the 
$\Omega$-effect (blue)) is the formation of sunspots on the surface of the Sun
from the rise of floating toroidal flux tubes from the base of the convection
zone. The magnetic fields of these sunspots closest to the equator in each
hemisphere diffuse and join, and the field due to the spots closer to the
poles has a polarity opposite to the current that initiates rotation of the
polarity. The newly formed polar magnetic flux is transported by the meridional
flow to deeper layers of the convection zone, thereby creating a new
large-scale poloidal field. Adapted from \cite{Sanchez2014}.
(b) BL~mechanism and holographic BL~mechanism as components of our
solar antidynamo model. Unlike the component of the solar dynamo model (a), the
BL~mechanism, which is predetermined by the fundamental holographic principle
of quantum gravity, and consequently, the formation of the thermomagnetic
EN~effect (see \cite{Spitzer1962,Spitzer2006,Rusov2015}),
emphasizes that this process is associated with the continuous transformation
of toroidal magnetic energy into poloidal magnetic energy
($T \rightarrow P$ transformation), but not vice versa ($P \rightarrow T$). This
means that the holographic
BL~mechanism is the main process of regeneration of the primary toroidal field
in the tachocline, and thus, the formation of floating toroidal magnetic flux
tubes at the base of the convective zone, which then rise to the surface of the
Sun. The joint connection between the poloidal and toroidal magnetic fields is
the result of the formation of the so-called meridional magnetic field, which
goes to the pole in the near-surface layer, and to the equator at the base of
the convection zone.
}
\label{fig-solar-dynamos}
\end{figure*}

The problem is devoted to the study of physics and the magnitude of the 
toroidal magnetic field in the tachocline. It is known that the radiation zone
of the Sun rotates approximately as a solid body, and the convection zone has
a differential rotation. This leads to the formation of a very strong shear 
layer between these two zones, which is called the tachocline. This gives rise
to a fairly simple and at the same time very complicated question: ``What is the
nature and the consequence of the existence of tachoclines on the Sun or other
stars?''

We propose a very unexpected answer that the existence of a two-dimensional
surface of the tachocline in solar interior is the manifestation of the 
holographic principle in the Universe and, therefore, on the Sun. It is important
to note that the holographic theory correlates the physical laws that act in 
some volume with the laws that act on the surface that limits this volume. The physics
at the boundary is represented by quantum particles that have ``colored'' 
charges and interact almost like quarks and gluons in standard particle
physics. The laws in the volume are a kind of string theory that includes the
force of gravity (see \cite{Maldacena2005}), which is difficult to describe in
terms of quantum mechanics. The main result of the holographic principle is the
fact that surface physics and physics in the volume are completely equivalent,
in spite of the completely different ways of describing them.

In most situations, the contradictory requirements of quantum mechanics and
general relativity are not a problem, because either quantum or gravitational
effects are so small that they can be neglected. However, with a strong
curvature of space-time, the quantum aspects of gravity become essential, and 
the conflict between the theory of gravity and quantum mechanics should disappear 
(see \cite{tHooft1993,Susskind1995,Maldacena1999,Hanada2014}). To create a
large curvature of space-time, a very large mass or density is
required. Some physicists believe that even the Sun is incapable of distorting
space-time so that the manifestations of quantum gravity effects become obvious.

Unlike other physicists, we believe that the fundamental holographic principle 
of quantum gravity predicts the experimental possibility of observational
measurements of magnetic fields between the two-dimensional surface of the
tachocline and the three-dimensional volume of the core on compact objects --
our Sun, magnetic white dwarfs, accreting neutron stars and BHs.

This is due to the fact that in addition to the postulate of the Hooft's brick
wall \citep{tHooft1993}, by now there is a lot of evidence that the physics of 
anti-de Sitter/conformal field theory correspondence (AdS/CFT) is correct.
In addition, there are some derivations based on physical arguments.

Below we show the connection between the black hole in AdS/CFT and the
tachocline of our Sun.

In theoretical physics the AdS/CFT correspondence
(\cite{Maldacena1998,Maldacena1999}), sometimes referred to as ``the Maldacena
duality'' or ``the gauge/gravity duality'' (see 
\cite{Gubser1998,Witten1998,Maldacena2015}), represents a hypothetical
connection between two different types of physical theories -- the gravitation
theory and the quantum field theory. The AdS spaces used in the quantum
gravitation theory formulated in terms of the string theory or M-theory
\citep{Becker2008} are on the one hand. The conformal field theories (CFTs),
which are the quantum field theories including those similar to the Yang-Mills
theories (see e.g. \cite{tHooft1972,tHooft2005}) describing the elementary
particles, are on the other hand. Such equivalence is an example of the
holographic duality -- a representation of the 3D space on the 2D photographic
emulsion (see \cite{tHooft1993,Susskind1995,Hanada2014}). Thus, although these
two types of theories seem to be very different, they are mathematically
identical \citep{Maldacena1999,Gubser1998,Witten1998}).

For example, according to \cite{Hanada2014}, a remarkable property of the
holographic principle of quantum gravity was obtained by Monte Carlo
simulation of the dual gauge theory for the parameters corresponding to a black
hole, which is destabilized due to the effects of quantum gravity.
The major results of \cite{Hanada2014} for the black hole mass are exactly
consistent with the prediction by \cite{Hyakutake2014} obtained from
independent calculations in gravity theory at the leading order of quantum
corrections. \cite{Hanada2014} thus obtained a quantitative evidence that the
dual gauge theory proposed in superstring theory provides a correct description
of a black hole, including the effects of quantum gravity.

Thus, we understand the connection between the black hole in AdS/CFT and the
tachocline of our Sun.

It means that, using the thermomagnetic EN~effect, a simple estimate of the
magnetic pressure of an ideal gas in the tachocline of e.g. the Sun can
indirectly prove that by using the holographic principle of quantum gravity,
the repelling toroidal magnetic field of the tachocline exactly ``neutralizes''
the magnetic field in the Sun's core

\begin{equation}
B_{tacho}^{Sun} = 4.1 \cdot 10^7 ~G = -B_{core}^{Sun}
\label{eq07-neutralize}
\end{equation}

\noindent
(see Eq.~(\ref{eq06-16}) and Fig.~\ref{fig-R-MagField}) 
where the projections of the magnetic fields of the
tachocline and the core have the equal value but the opposite directions.

Such strong magnetic fields in a tachocline of e.g. the Earth, magnetic white
dwarfs, accreting neutron stars and BHs can predict the exact ``neutralization''
of the magnetic field in the core of these stars and in a black hole (see
Sect.~4 in \cite{RusovArxiv2019}).

Let us note another important Babcock-Leighton (BL) mechanism
(see Fig.~\ref{fig-solar-dynamos}). On the one hand, we mark out the holographic
BL mechanism (Fig.~\ref{fig-solar-dynamos}b), which we often refer to as the
holographic antidynamo mechanism, caused by a remarkable example of the
Cowling antidynamo theorem. This theorem states that no axisymmetric magnetic
field can be maintained through the self-sustaining action of the dynamo by
means of an axially symmetric current~\citep{Cowling1933}. On the other hand,
the holographic BL~mechanism (as a component of our solar anti-dynamo model)
follows our example of a thermomagnetic EN~effect, or the so-called solar
holographic antidynamo, in which the poloidal field originates directly from
the toroidal field, as shown in Fig.~\ref{fig-solar-dynamos}b, but not vice
versa (see Fig.~\ref{fig-solar-dynamos}a).

\bibliographystyle{elsarticle-num}

\bibliography{Rusov-AxionSunLuminosity}

\end{document}